\pdfoutput=1
\documentclass[onecolumn]{aa} 

\usepackage{graphicx}
\usepackage{txfonts}
\usepackage{ulem}
\usepackage{amstext}
\usepackage{color}

\begin{document}

   \title{Magnetic topology of the north solar pole}

   \subtitle{}

   \author{A. Pastor Yabar
          \inst{1,2,3}
          \and
          M. J. Mart\'{\i}nez Gonz\'alez
          \inst{1,2}
          \and
          M. Collados
          \inst{1,2}
          }

   \institute{Instituto de Astrof\'{\i}sica de Canarias,
             C. V\'{\i}a L\'actea s/n., 38205 La Laguna, Tenerife, Spain\\
             \email{apastor@iac.es}
         \and
             Departamento de Astrof\'{\i}sica de Canarias, Universidad de La Laguna,
             Avda Astrof\'{\i}sico S\'anchez s/n., 38205, La Laguna, Tenerife, Spain
         \and
             Kiepenheuer-Institut f\"ur Sonnenphysik,
             Sch\"oneckstr. 6, 79104 Freiburg, Germany
             }

   \date{Received XX, XXXX; accepted XX, XXXX}

 
  \abstract
   {The magnetism at the poles is similar to that of the quiet Sun in the sense that no active regions are present there. However, the polar quiet Sun is somewhat different from that at the activity belt as it has a global polarity that is clearly modulated by the solar cycle. We study the polar magnetism near an activity maximum when these regions change their polarity, from which
it is expected that its magnetism should be less affected by the global field. To fully characterise the magnetic field vector, we use deep full Stokes polarimetric observations of the 15648.5 {\AA} and 15652.8 {\AA} Fe{\sc I} lines. We observe the north pole as well as a quiet region at disc centre to compare their field distributions. In order to calibrate the projection effects, we observe an additional quiet region at the east limb. We find that the two limb datasets share similar magnetic field vector distributions. This means that close to a maximum, the poles look like typical limb, quiet-Sun regions. However, the magnetic field distributions at the limbs are different from the distribution inferred at disc centre. At the limbs, we infer a new population of magnetic fields with relatively strong intensities ($\sim$ 600-800 G), inclined by $\sim$30$^\circ$ with respect to the line of sight, and with an azimuth aligned with the solar disc radial direction. This line-of-sight orientation interpreted as a single magnetic field gives rise to non-vertical fields in the local reference frame and aligned towards disc centre. This peculiar topology is very unlikely for such strong fields according to theoretical considerations. We propose that this new population at the limbs is due to the observation of unresolved magnetic loops as seen close to the limb. These loops have typical granular sizes as measured in the disc centre. At the limbs, where the spatial resolution decreases, we observe them spatially unresolved, which explains the new population of magnetic fields that is inferred. This is the first (indirect) evidence of small-scale magnetic loops outside the disc centre and would imply that these small-scale structures are ubiquitous on the entire solar surface. This result has profound implications for the energetics not only of the photosphere, but also of the outer layers since these loops have been reported to reach the chromosphere and the low corona.}


   \maketitle
%

\section{Introduction}

Solar polar caps are the areas that cover heliographic latitudes above 60$^{\circ}$, which means that these are quiet areas, devoid of active regions (hereafter ARs). The research on polar photospheric magnetic fields started in the mid-1950s and was conducted using low-resolution magnetograms \citep{babcock1955}. Babcock and
Babcock detected for the first time an average line-of-sight (LOS) magnetic field component of $<B_{LOS}>\approx1$G over the polar caps. This LOS magnetic flux was found to have a dominant polarity of opposite sign at the north and south poles. Using the same technique, the same group later found that the polar magnetic field was modulated by the activity cycle, which reaches maximum values during sunspot activity minima and reverts the polarity in periods with a maximum occurrence of sunspots \citep{babcock1959}. This meant that the polar magnetism was half a cycle (5-6 years) out of phase from the activity belt \citep[latitudes below 40$^{\circ}$; see e.g. Sect. 3.6 of][]{hathaway2015}. Constrained by these observational facts, the first dynamo theory was proposed by \cite{babcock1961} and \cite{leighton1964}, explaining this behaviour in the cycle context. In this theory, part of the emerging flux at the activity belt is dragged to the polar regions. This flux advection takes place so that a dominant polarity is driven to the polar regions (PRs), and it has opposite sign for each PR. Finally, the flux that is piled up in each PR feeds the next solar cycle. The PR magnetism was recognised as a significant factor in understanding the solar cycle magnetism.

Triggered by these early results, the polar caps magnetism has been the subject of deep and continuous study. The magnetism in these regions is a key ingredient for the study of the axisymmetric component of the global solar magnetism \citep{knaack2005} and for the dynamo theory \citep{cameron2015}. In contrast to the activity belt, where active regions appear with balanced polarities, in the PR, most of the strong ($kG$) magnetic structures share the same magnetic polarity. These vertical $kG$ fields are assumed to be the roots of the open field lines of the Sun. Through these open field lines, the fast solar wind is thought to be funnelled to the heliosphere \citep{krieger1973}.

Most of the research of the polar region photospheric magnetism has been conducted by means of LOS magnetograms. A deeper characterisation of the solar pole magnetism is possible by means of magnetic field vector studies. Such works require full Stokes spectropolarimetric observations. For the quiet-Sun (hereinafter QS) magnetism, which is characterised by very low polarisation signals and small-scale structures, observations with high resolution and high sensitivity are mandatory. This last point has prevented any topological study until the past decade. Because polar observations take place at high heliocentric angles with associated complications in the posterior analysis of the magnetic topology (foreshortening, change on the viewing angle, limb effects), only a few works have studied the topology of PR magnetic structures. Observations
with the spectropolarimeter \citep[SP,][]{lites2001} of the Solar Optical Telescope \citep[SOT,][]{tsuneta2008, suematsu2008, ichimoto2008, shimizu2008} on board the Hinode satellite \citep{kosugi2007} have been particularly fruitful for the study of the polar magnetism \citep{tsuneta2008b, ito2010, jin2011, shiota2012, kaithakkal2013, kaithakkal2015, quinteronoda2016}.

\begin{figure*}
  \centering
    \includegraphics[width=1.\textwidth]{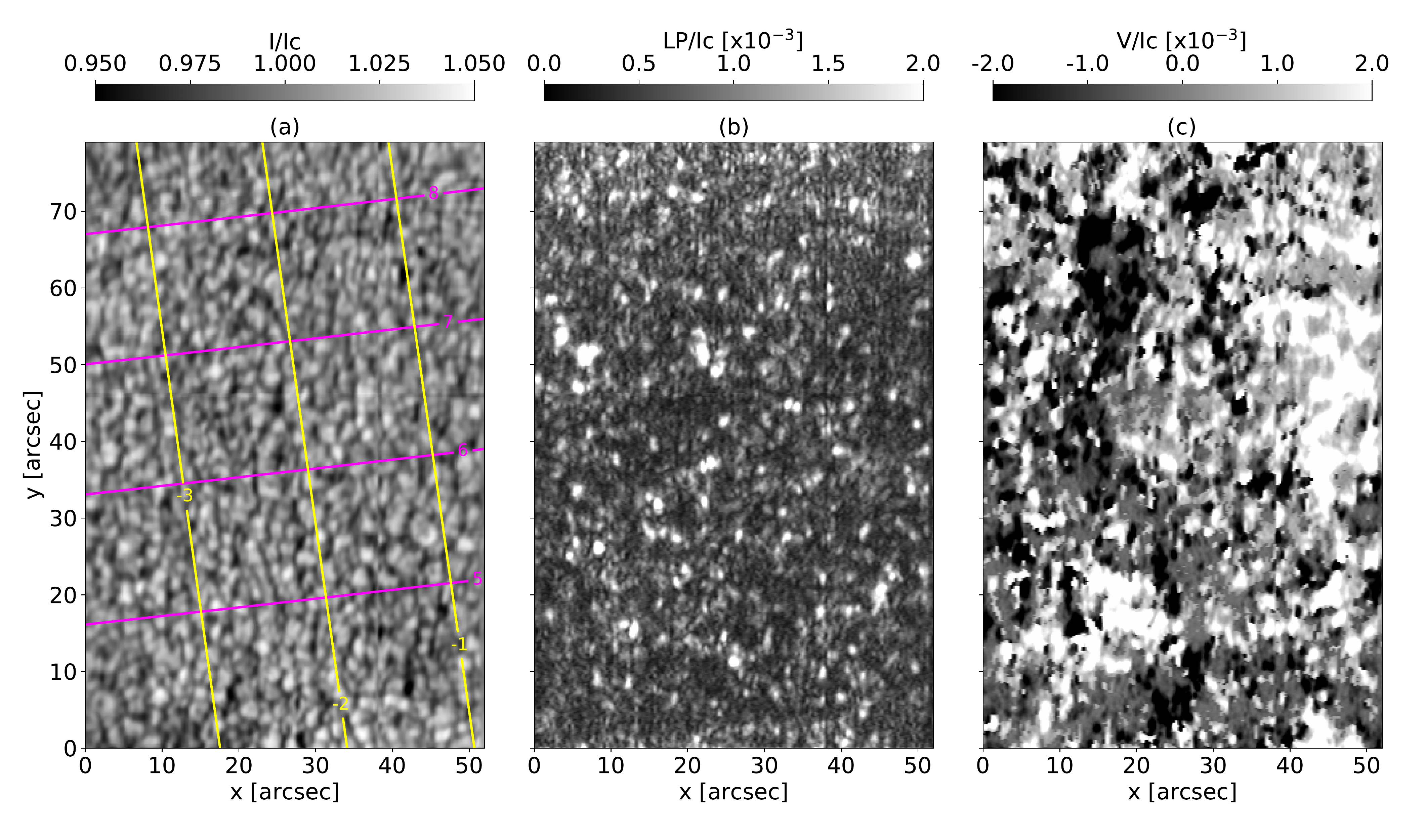}
  \caption{Intensity and polarisation maps for the disc centre dataset. Panel {\it a}: Continuum intensity map normalised to the average continuum intensity. Magenta and yellow lines mark the solar latitude and longitude, respectively. The x- and y-axes show the observed distance on the solar disc in arcseconds. $I_{c}$ is the average continuum intensity of disc centre quiet Sun. Panel {\it b}: Map of the maximum linear polarisation signals at the blue wing of the 15648.5{\AA} spectral line (given by $\sqrt{Q_{\lambda}^2+U_{\lambda}^2}$). Panel {\it c}: Map of the maximum amplitude Stokes V at the blue wing of the 15648.5{\AA} spectral line.}
  \label{fig:fovmaps1}
\end{figure*}

\begin{figure*}
  \centering
    \includegraphics[width=0.8\textwidth]{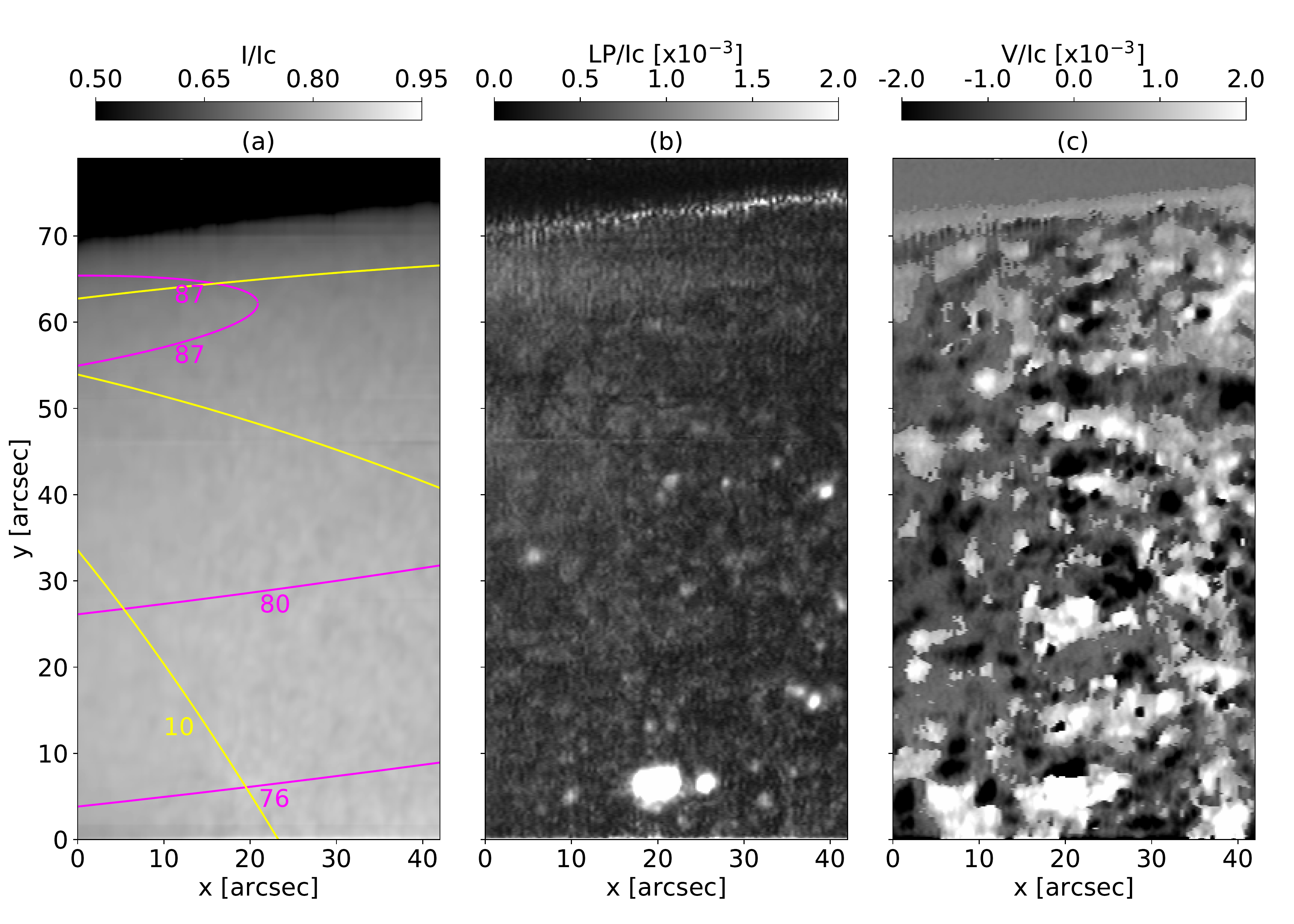}
  \caption{Same as Fig. \ref{fig:fovmaps1} for the north limb dataset.}
  \label{fig:fovmaps2}
\end{figure*}

\begin{figure*}
  \centering
    \includegraphics[width=1.\textwidth]{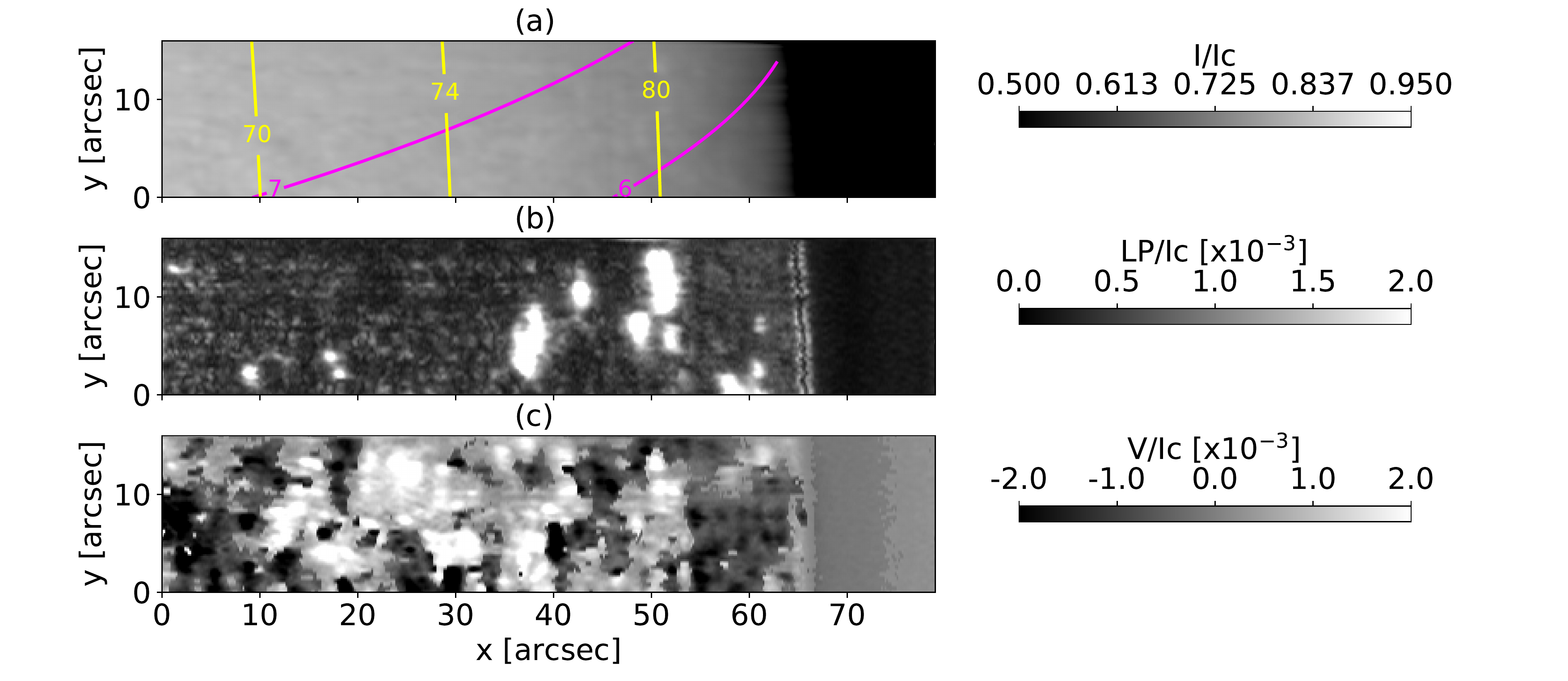}
  \caption{Same as Fig. \ref{fig:fovmaps1} for the west limb sample.}
  \label{fig:fovmaps3}
\end{figure*}

\cite{tsuneta2008b} confirmed that the magnetism at the PRs is concentrated in small ($1^{\prime\prime}\times1^{\prime\prime}$ up to $5^{\prime\prime}\times5^{\prime\prime}$) magnetic patches that are scattered throughout the region. These patches are divided into two main groups: vertical and horizontal fields. These authors found that horizontal magnetic fields presented strengths of the order of $hG$ and had balanced polarities. In contrast, vertical magnetic fields were characterised by $kG$ strengths and a dominant polarity that coincided with that of the PR.

\cite{ito2010} compared the magnetism at the PR and that of a QS region at the east limb, following the same strategy as \cite{tsuneta2008b}: separating the field populations into vertical and horizontal. They found that the horizontal component showed a similar statistical behaviour at the polar and equatorial QS regions. A different property was found for the vertical component at the PR. Close to the pole, this magnetic component showed more patches with vertical fields. In addition, some of these patches were also larger and carried more magnetic flux than those found at the east limb. The authors also obtained that the vertical fields presented a dominant negative polarity, which means that of the overall PR. In contrast, in the east limb QS, they found a polarity balance even for the population of vertical kG fields. \cite{jin2011} performed a similar study and additionally studied the magnetism of the QS within a coronal hole. They confirmed the previous results by \cite{ito2010} and also found that the PR and a coronal hole showed a similar fraction ($\text{one-third}$) of vertical patches open to the interplanetary medium. \cite{shiota2012} again followed the classification of the fields into vertical and horizontal and studied their behaviour between 2008 to 2012 for the PRs and equatorial QS. These authors found that only the largest magnetic flux concentrations belonging to the vertical component carried the polarity reversal of the PR, while the remaining vertical fields as well as the horizontal component behaved independent of the cycle phase. 

A known observational feature related to the magnetism is faculae \citep{homann1997,okunev2004,blancorodriguez2007}. These features are magnetic concentrations brighter than the surroundings in white light, and in the polar regions they are called polar faculae (PF). The number of these features in the PR varies with the solar cycle \citep{sheeley1964}. They are more abundant close to the cycle minimum, when PRs exhibit their strongest averaged magnetic signals. \cite{kaithakkal2013} studied the relation of these features to the vertical magnetic patches. They found that most of the largest magnetic flux concentrations hosted faculae. A deeper insight into the thermodynamic and magnetic properties of polar magnetic structures was gained by \cite{quinteronoda2016}. They studied the PF height dependence of different physical properties
and found that PF are characterised by vertical fields with hot plasma and with a mixture of positive and negative velocities at the bottom that gradually decreased with height, showing null values at the top layers.

All the previous works used SP/SOT data at visible wavelengths. \cite{blancorodriguez2010} exploited the near-infrared domain using full spectropolarimetric data obtained with the Tenerife Infrared Polarimeter-II \citep[TIPII,][]{collados2007} at the Vacuum Tower Telescope \citep[VTT,][]{schroeter1985}. They studied both north and south PRs and found that the strongest magnetic fields were associated with PF and characterised by vertical magnetic fields. They also found a bimodal behaviour in the field strengths of these structures, with two types of faculae. Some of them had strong magnetic fields (ranging from 900 to 1500G) and were compatible with inclinations vertical to the surface. These fields presented a dominant polarity at each PR, which was of the same polarity as each PR averaged polarity. They also found other PF with weaker fields (below 900G) with no preferred orientation. 

We aim to continue deepening our understanding of the photospheric magnetism in PRs, in particular concerning the magnetic topology of polar fields. To do so, we perform a topological study  close to solar maximum, which has been done by very few studies \citep{shiota2012}. Here we use full spectropolarimetric observations of the near-IR spectral lines at 1.565 $\mu$m, as in \cite{blancorodriguez2010}, to reconstruct the vector magnetic field in the north PR and in the equatorial QS. Longer wavelengths provide a higher magnetic sensitivity and hence a more precise characterisation of the weakest magnetic fields. We describe the observations, reduction, and inversion methods in Sect. \ref{Sect:observation_analysis}, the results are presented in Sect. \ref{Sect:results}, and the conclusions are discussed in Sect. \ref{Sect:discussion_conclusions}.

\section{Observations and analysis}
\label{Sect:observation_analysis}

\begin{table*}
  \begin{center}
    \caption{Heliographic coordinates of the central point of each scan, and their spatial size. \label{tab:fovdetails}}
    \begin{tabular}{lccc}
      \hline\hline
        & Disc Centre & North Pole & West Limb \\
      Heliographic coordinates (x;y) & -35$.\!\!^{\prime\prime}$7; -16$.\!\!^{\prime\prime}$7 & 8$.\!\!^{\prime\prime}$2; 921$.\!\!^{\prime\prime}$3 & 926$.\!\!^{\prime\prime}$2; -79$.\!\!^{\prime\prime}$4 \\
      FOV & 78$.\!\!^{\prime\prime}58\times52.\!\!^{\prime\prime}$50 & 78$.\!\!^{\prime\prime}75\times42.\!\!^{\prime\prime}$00 & 78$.\!\!^{\prime\prime}93\times15.\!\!^{\prime\prime}$75 \\
    \hline
    \hline
    \end{tabular}
    \label{Tab:FOVs}
  \end{center}
\end{table*}

\subsection{Observations}

The data analysed here were taken on 14 September 2013, when the north pole could be optimally observed from the ground. In the solar cycle context, this corresponded to an epoch close to the maximum of activity, that is, to the moment when the reversal of the dominant polar polarity takes place.

The data were recorded using the Tenerife Infrared Polarimeter \citep[TIP II, ][]{collados2007} installed at the VTT at the {\it Observatorio del Teide} (Spain). The observed spectral region was centred at 1.565 $\mu$m and spanned 14 {\AA} with a spectral sampling of 14 m{\AA}. In this spectral range, two neutral iron spectral lines are located that are highly sensitive to magnetic fields: 15648.515 {\AA} and 15652.874 {\AA}, with a Land\'e factor of 3.0 and 1.48, respectively. The high Land\'e factor, together with the fact that the Zeeman sensitivity increases with wavelength, makes this pair of spectral lines an excellent candidate for studying the QS magnetism.

The polarimeter was attached to a spectrograph with a slit of 150 $\mu$m width. The width of the slit on the sky was 0.$^{\!\!\prime\prime}$68, and in order to perform the scan, the image was shifted perpendicular to the slit in steps of 0.$^{\!\!\prime\prime}$35. The slit width was slightly higher than the spatial resolution imposed by the diffraction of the telescope (0.$^{\!\!\prime\prime}$56). In principle, the element limiting the spatial resolution was the slit. However, the seeing conditions decreased the spatial resolution of the observations to about 1.$^{\!\!\prime\prime}$1, as determined from the Fourier power spectrum of the continuum intensity images.

We performed a deep observation with an effective exposure time of 20 seconds per slit position (giving a cadence of 23.7 seconds, including overheads). The strategy we followed for the observation was similar to that performed by \cite{ito2010, blancorodriguez2010} and \cite{jin2011}. In order to determine
whether the PR magnetism was intrinsically different from the QS magnetism at low latitudes, an additional QS region close to disc centre was observed. However, the different viewing angles of these two disc positions might induce differences in the observations and their posterior analysis. For complementarity, we observed an additional dataset at one of the equatorial limbs. When we
assume that the QS at disc centre and equatorial limb sample the same statistical magnetic distribution, the comparison of these two datasets allows estimating the projection effects when observing close to the limb.

The projection effects include two main contributions. First, there are purely geometrical effects (e.g., a vertical magnetic field vector at disc centre is seen as perpendicular to the LOS at the limb). Second, limb effects include 1) limb darkening, 2) spectral lines close to the limb that come from higher layers than when observing at disc centre, and 3) changes in the radiative transfer due to the different physical magnitude stratifications in the optical path. These issues may cause different polarisation signals when observing the magnetism at disc centre and at the limb, even when both regions have statistically the same properties.

To study the QS, we avoided active areas using the CaII K slit-jaw images, where bright regions are associated with strong flux concentrations. Table \ref{Tab:FOVs} shows the position and size of each observed field of view (FOV). In Figs. \ref{fig:fovmaps1}, \ref{fig:fovmaps2}, and \ref{fig:fovmaps3}, we show the spatial maps of the continuum intensity, the linear polarisation, and the circular polarisation for the three datasets. The linear polarisation was measured by taking the maximum value of the quantity $\sqrt{Q^2+U^2} / I_\mathrm{c}$ around the 15648.5{\AA} spectral line, where $Q$ and $U$ refer to the Stokes parameters for the linear polarisation, and $I_\mathrm{c}$ represents the local continuum intensity. The circular polarisation maps show the signed maximum of the blue component found for the 15648.5{\AA} spectral line. These maps show the strongly magnetised character of QS regions through the ubiquitous polarisation signals throughout
the FOV, even for linear polarisation signatures, which are usually more difficult to detect.

\subsection{Data reduction}

\begin{figure}
\centering
\includegraphics[width=0.5\textwidth]{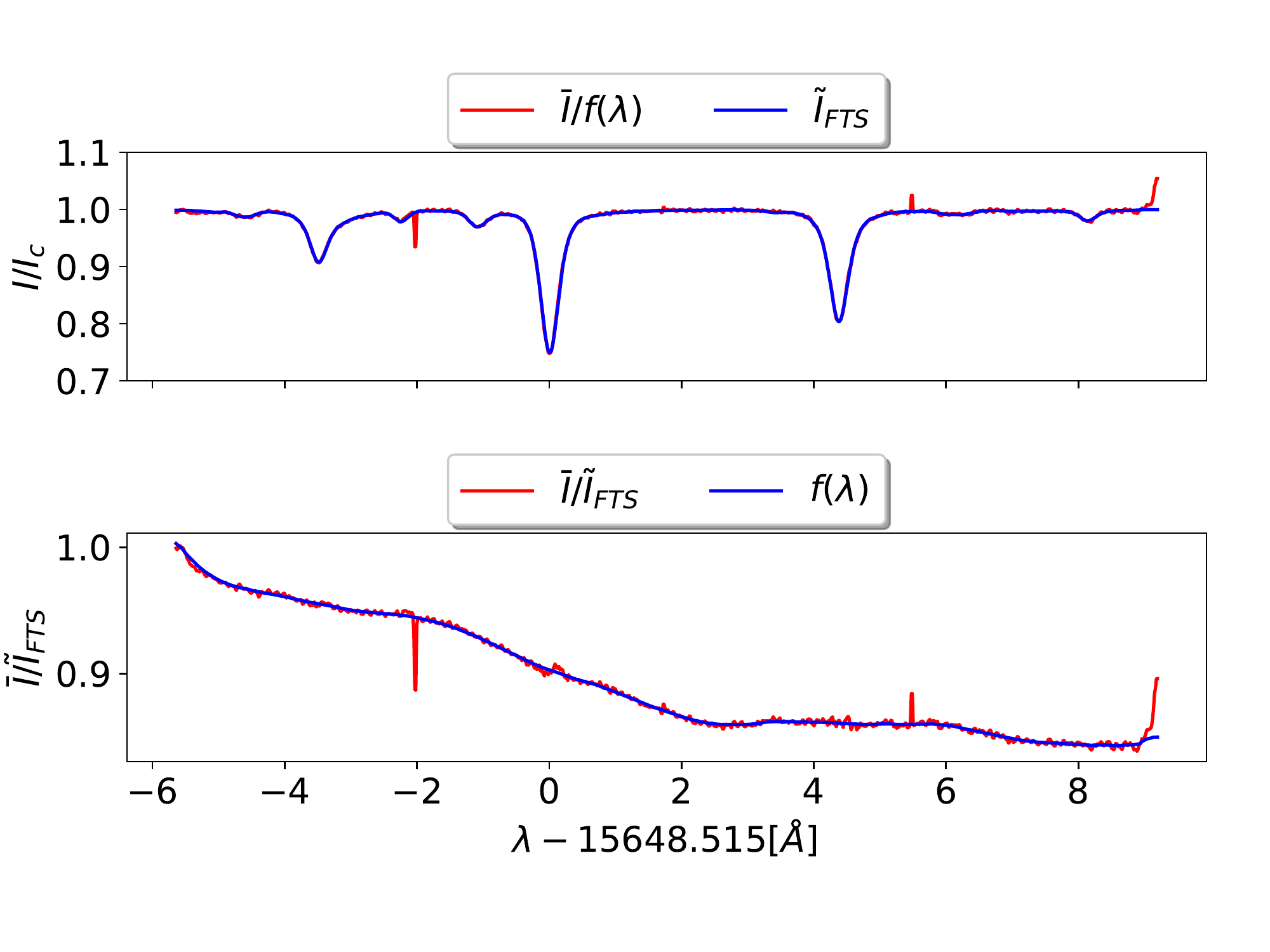}
\caption{Correction of the continuum intensity trends. Top panel: Comparison between the degraded FTS spectrum $\tilde{I}_{FTS}(\lambda)$ (blue) and the mean intensity profile observed, $\bar{I}$ corrected for the continuum oscillations $f(\lambda)$ (red). Bottom: Ratio between $\bar{I}$ and $\tilde{I}_{FTS}(\lambda)$ (red) and the fitted function $f(\lambda)$ (blue).
        }
\label{Fig:tipcontinuumcorrection}
\end{figure}

\begin{figure*}
  \includegraphics[width=\textwidth]{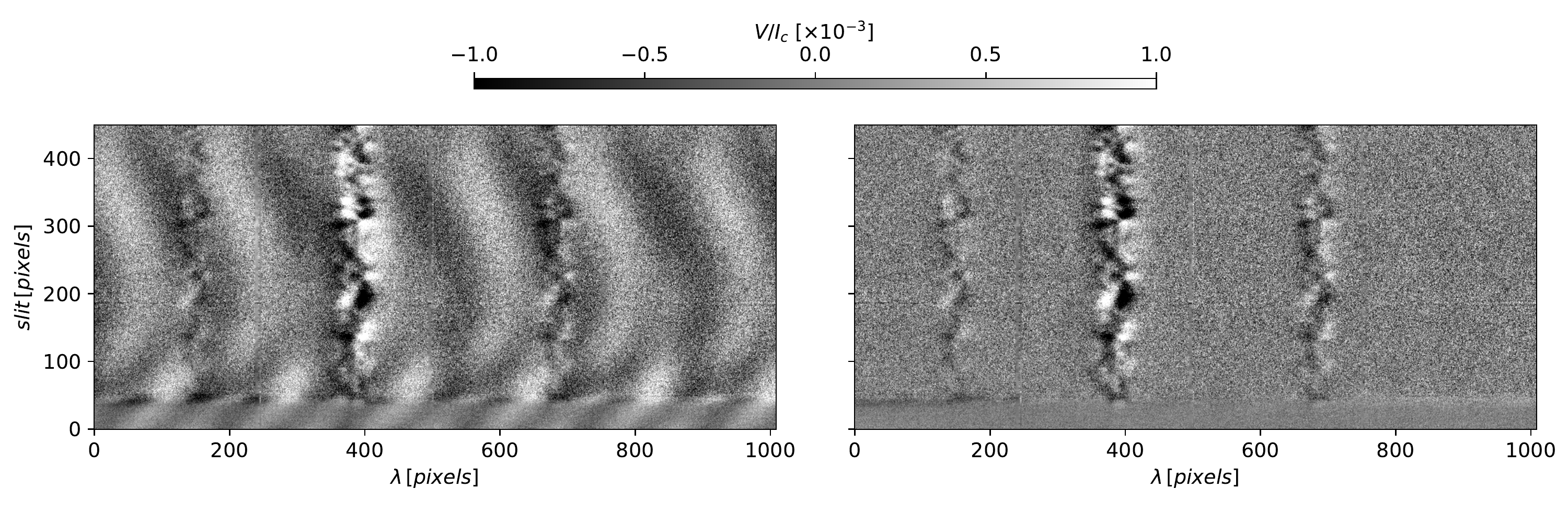}
    \caption{Polarimetric interference fringe correction. Left panel: Stokes V spectral map for slit position 10 for a north PR dataset before correction. Right panel: Same map after removing polarimetric interference fringes.}
  \label{Fig:filteredspectralmaps}
\end{figure*}

Instrument developers supply software in order to process raw data to ready-to-analyse data products. This software provides the dark current subtraction, the flat-field correction, the bad-pixel correction, the demodulation of the data, and the instrumental cross-talk removal. However, since QS signals are typically very low, we need to push the polarimetric accuracy to the limit. Hence, after the standard reduction, we still had to process the data to remove some features that were at the noise level or slightly higher. These features were mainly residuals of the flat-field correction and cross-talk between the Stokes parameters. In our case, since we compared three different datasets at different positions of the solar disc, we also needed to refer all signals to a common reference frame. We took as the reference for intensities the average continuum intensity at disc centre,
$I_\mathrm{c}$.

Our observational strategy comprised two flat-field measurements at disc centre, one just before the science observation, and another one at the end of it. Since mainly the interference fringes of the detector change during the observation, having two flat-field measurements allows the standard reduction routine to interpolate the flat-field correction to the time of the science observation. This procedure works very well in general. In our case, however, residual errors of the standard flat-field correction prevented us from detecting the weakest polarimetric signals of the QS. We first corrected for residual intensity trends (mostly seen in the continuum of the spectrum) that persisted after flat-fielding.

These light trends were estimated with a polynomial fit to the mean intensity profile of the flat-field after being reduced with the standard software as if they were science data ($\bar{I}$). The spectral lines in $\bar{I}$ might influence this fit. Therefore we chose to use the reduced flat-field, since these images were taken at disc centre, allowing the removal of the spectral lines using spectra of the Fourier Transform Spectrometer (FTS, \citealt{livingston1991}) as a reference. The FTS is a spectrum with a  very high signal-to-noise ratio with no significant spectral broadening, no spectral stray light, and with a very accurate continuum level. Using the flat-field to compute the spurious light trends to correct all the data sets, we also normalised them to the common reference luminosity of the disc centre.

Following \cite{allendeprieto2004}, the intensity spectrum of the FTS ($I_{FTS}$) was modified to match our $\bar{I}$ using the following expression:

\begin{equation*}
\tilde{I}_{FTS}(\lambda)=\alpha+(1-\alpha)\Gamma(\sigma)\star I_{FTS}(\lambda),
\end{equation*}

\noindent where $\alpha$ is a white-light stray light factor. To decrease the spectral resolution of the FTS to that of our observations, we convolved the $I_{FTS}(\lambda)$ with a Gaussian function $\Gamma(\sigma)$ centred at zero and with a standard deviation $\sigma$. However, the values of $\alpha$ and $\sigma$ were affected by the spurious trends, which we wished to remove, present in $\bar{I}$. The trend estimation again depended on subtracting the spectral lines. This is a coupled inversion problem, and we followed an iterative process in order to retrieve the $\sigma$ and $\alpha$ parameters and the light trends, as described
below. 
\begin{enumerate}
\item We determined the best fit of the $\tilde{I}_{FTS}(\lambda)$ to $\bar{I}/f(\lambda)$, where $f(\lambda)$ is an initial guess of the continuuem fluctuations. In the first iteration, $f(\lambda)$ is given by a linear fit to $\bar{I}$. In this step, $\alpha$ and $\sigma$ are free parameters,
\item The FTS estimate degraded to our instrumental setup ($\tilde{I}_{FTS}(\lambda)$) allows a spectral line subtraction from the mean flat-field spectrum $\bar{I}$ by defining a transmission function. In this way, we can better estimate the light trends in our data by fitting $\bar{I}/\tilde{I}_{FTS}(\lambda)$ with a polynomial of up to order 20. $f(\lambda)$ is set to the best fit of these polynomial functions (as deduced from the minimum $\chi^2$).
\item This new estimate $f(\lambda)$ feeds the next cycle. When convergence is achieved, $f(\lambda)$ contains the intensity trend to correct all the Stokes parameters. As a result, $\sigma$ is representative of the spectral resolution, and $\alpha$ gives an estimate of the amount of white-light contamination.
\end{enumerate}

Convergence is reached when the reduced $\chi^2$ between $\bar{I}/f(\lambda)$ and $\tilde{I}_{FTS}(\lambda)$ is below 0.01 or more than 100 cycles have evaluated the $\chi^2$. Figure \ref{Fig:tipcontinuumcorrection} shows an example of the result. The upper panel compares $\bar{I}/f(\lambda)$ and $\tilde{I}_{FTS}(\lambda)$, plotted as red and blue lines, respectively. The lower panel depicts $\bar{I}/\tilde{I}_{FTS}(\lambda)$ and $f(\lambda)$ in red and blue, respectively. The correction $f(\lambda)$ is applied to each of the Stokes parameters. The values retrieved for this dataset are $\alpha=4.23\%$ and \mbox{$\sigma=53.47$ m\AA.}

As a sub-product of this correction, we calibrated the data wavelength. After the final fit, we obtained the atlas wavelengths, and hence the wavelength calibration of the data. Since the atlas uses an absolute wavelength calibration, our wavelength calibration is also absolute. The spectral sampling of the data is $14.7$ m{\AA}/px.

Interference fringes persist in the polarised spectra above the noise level (see the left panel of Fig. \ref{Fig:filteredspectralmaps}). We computed and corrected the polarised fringes at each individual pixel and each Stokes parameter. To estimate the fringes, we first fit the profiles with a pair of sinus and cosinus of the
given periods between 1 to 4 {\AA}. This fitting was done avoiding the spectral line regions. Then, only the sinus-cosinus pair with the period that corrects the interference fringes most was applied. An example of the result of this fringe subtraction is presented in the right panel of Fig. \ref{Fig:filteredspectralmaps}.

\begin{table*}
  \caption{Atomic parameters of the various spectral lines. From left to right: Atomic element ({\it El}), ionisation state of the atom ({\it ion}), wavelength of the transition in angstrom ($\lambda$), excitation potential of the lower level ({\it E}), oscillator strength ({\it log(gf)}), terms of the lower and upper levels, Barklem velocity exponential ($\alpha$), and Barklem cross-section in units of cm$^{-2}$ ($\sigma$).}
  \label{Tab:ATOM}
  \begin{center}
    \begin{tabular}{cccccccccc}
      \hline
      \hline
        El & ion & $\lambda$ & E & log(gf) & $^{2S+1}L_{J}\,(l)$ & $^{2S+1}L_{J}\,(u)$ & $\alpha$ & $\sigma$ \\
        - & - & \AA\ & eV & - & - & - & - & cm$^{-2}$ \\
      \hline
Fe & {\sc I} & 15648.515 & 5.426 & -0.669 & $^7D_{1.0}$ & $^7D_{1.0}$ & 0.229 & 2.744e-14 \\
Fe & {\sc I} & 15652.874 & 6.246 & -0.095 & $^7D_{5.0}$ & $^7k_{4.0}$ & 0.330 & 3.992e-14 \\
      \hline
    \end{tabular}
  \end{center}
\end{table*}

The standard cross-talk correction \citep{schlichenmaier2002} assumes that the deviation of Stokes $Q$, $U,$ and $V$ continua from zero is a contamination from Stokes $I$ (in the near-infrared scattering polarisation of the continuum is negligible). To correct the cross-talk of the intensity to the polarisation parameters, as much Stokes $I$ as needed is subtracted for Stokes $Q$, $U,$ and $V$ to have polarimetric continua at the zero level. The cross-talk from Stokes $V$ to $Q$ and $U$ is computed with a least-squares fitting in the pixels where the Stokes $V$ signal is stronger than $Q$ or $U$. The mean value of the least-squares coefficients is assumed as the amount of cross-talk from circular to linear polarisation. The crosstalk from circular to linear is corrected by removing from $Q$ and $U$ the amount of $V$ given by the determined cross-talk coefficient. This automatic procedure works accurately for on-disc data, leaving cross-talk residua well below the noise level. However, for limb data, it typically leaves a residual cross-talk between the different Stokes parameters that is higher than the noise level. The main reason is that the strong intensity variation at the solar limb creates strong polarisation signals
(although they vary with seeing conditions) that can reduce the accuracy  with which the cross-talk coefficients are determined. We modified the standard routine by \cite{schlichenmaier2002} by calculating the cross-talk coefficients with profiles only inside the solar disc.

The last step performed before we analysed the data consists of applying a technique to reduce the uncorrelated noise. This step was done using the principal component analysis \citep[PCA, ][]{loeve1955}. This technique decomposes a set of observations into an orthonormal basis of vectors. This basis is able to collect most of the correlated variance in the first eigenvectors. This technique can be applied to spectropolarimetric data in order to reduce uncorrelated noise of the Stokes profiles \citep{martinezgonzalez2008}. To do so, we performed the PCA decomposition of each dataset and recovered the Stokes profiles with the first 40 eigenvectors for each Stokes parameter. This number is larger than the usual number of profiles used for a PCA decomposition \citep{ruizcobo2013,quinteronoda2015}. This number of eigenvectors ensures that most of the information encoded in the profiles is preserved even at the expense of also preserving some amount of noise. We chose this number of eigenvectors by considering the eigencoefficient maps and preserving all the eigenvectors until these maps became noise. After the reduction process, the final polarisation sensitivity is $\sigma_n^{Q,U,V}/I_{c}=1\times10^{-4}$, where $I_{c}$ is the average continuum intensity at disc centre. We considered a positive detection of polarisation signatures when any of the Stokes parameters presented at least three consecutive wavelengths higher than a 5$\sigma_n$ criterion. Below this criterion, more than 70\% of the observed FOVs show a polarimetric detection (either in Stokes $Q$, $U,$ or $V$) in any of the three FOVs.

\subsection{Inversion}

\begin{table}
  \caption{Minimum and maximum values for the random initialisation of the various model atmospheric parameters.}
  \label{Tab:rangevalues}
  \begin{center}
    \begin{tabular}{lccc}
      \hline
      \hline
      Parameter & min & max & units\\
      \hline
      B & 0 & 2000 & G \\
      $\theta$ & 0 & 180 & $^{\circ}$ \\
      $\chi$ & 0 & 180 & $^{\circ}$\\
      v$_{LOS}^{mag}$ & -5 & 5 & km/s\\
      v$_{mic}^{mag}$ & 0 & 1 & km/s\\
      v$_{LOS}^{no-mag}$ & -5 & 5 & km/s \\
      v$_{mic}^{no-mag}$ & 0 & 1 & km/s \\
      \hline
    \end{tabular}
  \end{center}
\end{table}

The inversion of the data allows us to recover the physical parameters of the medium where the light was emitted from the information contained in the Stokes profiles. For the radiative transfer problem, some codes have been developed. In particular, we used the code called Stokes inversion based on response functions \citep[][SIR]{ruizcobo1992}. This code is based on the assumption of local thermodynamic equilibrium conditions to solve the radiative transfer equation for polarised light. In this situation, the scattering processes are decoupled from the radiation transfer problem since the atomic level populations and the ionisation states are given by the local conditions through the Saha and Boltzmann equations. In addition, the source function is given by the Planck function. This is a valid approximation for most spectral lines formed in the photosphere, as we used here.

The atomic parameters of the spectral lines used in the inversion are listed in Tab. \ref{Tab:ATOM}. The inversion was made by modelling each element by one magnetic and one non-magnetic atmosphere. We considered height-independent parameters except for the temperature. Departing from the HSRA \citep{gingerich1971} temperature stratification, we allowed perturbations in five independent nodes for each atmosphere. We additionally constrained the continuum value of the two atmospheres so that it was the same. The remaining parameters are the LOS velocity and microturbulent velocity for each atmosphere; the magnetic field strength, inclination, and azimuth of the magnetic atmosphere, and the macroturbulent velocity (which was chosen to be a common parameter for both atmospheres). All of them are considered to be constant in height, giving up to 18 free parameters. The inversion was performed several times (up to 50 successful inversions) for each pixel with different initial values for the magnetic and dynamic parameters. To do so, we set different random initial values for the magnetic field strength, inclination, and azimuth and the LOS and microturbulent velocities of the input models. These values are randomly within the ranges listed in Tab. \ref{Tab:rangevalues}. The result is given by the atmosphere model that best fits the observed full Stokes vector ("best" meaning the fit with the smallest chi-square).

\begin{figure}
\centering
\includegraphics[width=0.45\textwidth]{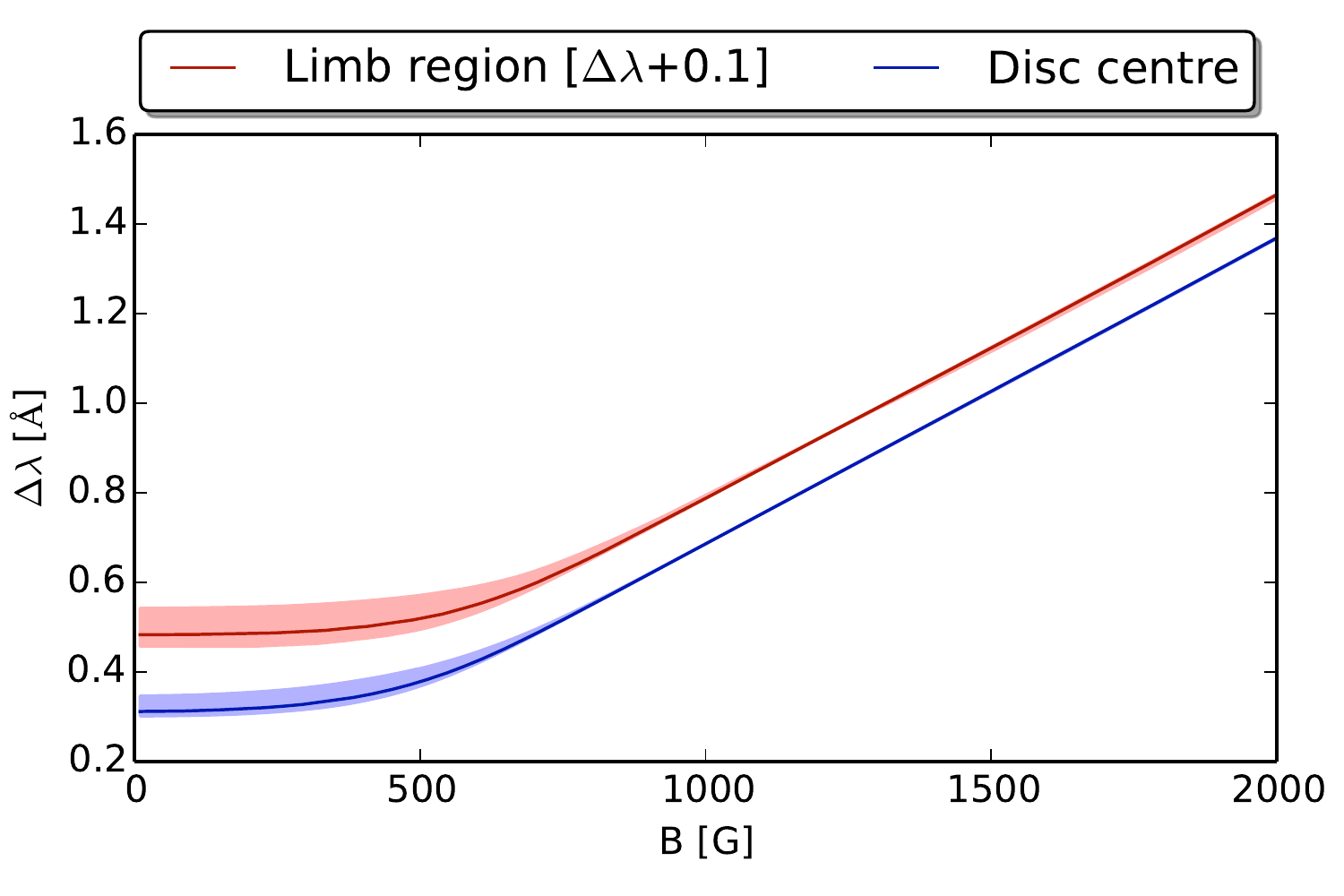}
\caption{Zeeman splitting dependence on magnetic field strength of the $\sigma$ components of the 15648.515{\AA} Fe {\sc I} spectral line for disc centre (blue) and limb regions (red). The splitting was measured for 1000 model atmospheres taken from the inversions. The coloured shadowed area stands for the splitting in between the 25th and 75th percentile. The coloured solid line represents the 50th percentile.}
\label{Fig:vttzeemansplitting}
\end{figure}

A well-known problem of inverting noisy data is that the model parameters are not always independent. In particular, when the magnetic field strength is not strong enough, the LOS inclination, the magnetic field strength, and the filling factor  are coupled
\citep{asensioramos2007}. In this case, these parameters cannot be determined independently, and only the LOS magnetic flux density is retrieved: $\phi=\alpha\,B\,\cos{\theta}$. This situation holds as long as the Zeeman splitting is below or of the order of the Doppler width. In order to avoid this ambiguity when analysing the magnetic field strength and inclination, we limited ourselves to those magnetic fields for which the Zeeman splitting is larger than the Doppler width. In order to set this limit, we took 1000 thermodynamic model atmospheres obtained from the disc centre and at the limb dataset inversions. For each atmosphere, we synthesised the Stokes profiles of the 15648.515 {\AA} spectral line (the line with the highest Land\'e factor) to increase magnetic fields from \mbox{0 to 2000 G}. We calculated the distance between the two lobes of the Stokes V profiles. Figure \ref{Fig:vttzeemansplitting} presents the result of this experiment. For the disc centre (blue),
the splitting between the two lobes is the same for the fields up to 400 G. When the field exceeds about 400 G, the distance between the lobes increases for some model atmospheres, and above 500 G, the splitting is linear with the magnetic field strength. In the following, the analysis of the disc centre data is restricted to pixels with magnetic fields higher than 500 G. For the limb data (red in Fig. \ref{Fig:vttzeemansplitting}), as spectral lines are broader the closer they are to the limb \citep{dravins1981}, the different atmospheric models cause this threshold value to
increase slightly, to about 600 G. This criterion is fulfilled by $\sim$16\%, $\sim$21\%, and $\sim$24\% of the polarimetric signals detected in the north, disc centre and west FOV, respectively. The criterion we used does not consider the linear polarisation signals. In the presence of Stokes $Q$ and $U$ signals, even below the mentioned magnetic field strength thresholds, it is possible to infer the magnetic field strength, inclination, and filling factor independently. We have found that linear polarisation signals at disc centre are rather weak and usually associated to irregular Stokes V profiles. In this case, it is very hard to unequivocally determine the magnetic field topology even in the presence of linear polarisation signals because the inversion becomes very sensitive to the model employed. For the limb datasets, this problem persisted. In addition there are some strong linear signals that already fulfil the previous magnetic strength criteria, so we did not need to add any other condition.

\section{Results}
\label{Sect:results}

\subsection{Disc centre QS}

\begin{figure*}
\centering
     \includegraphics[width=1.\textwidth]{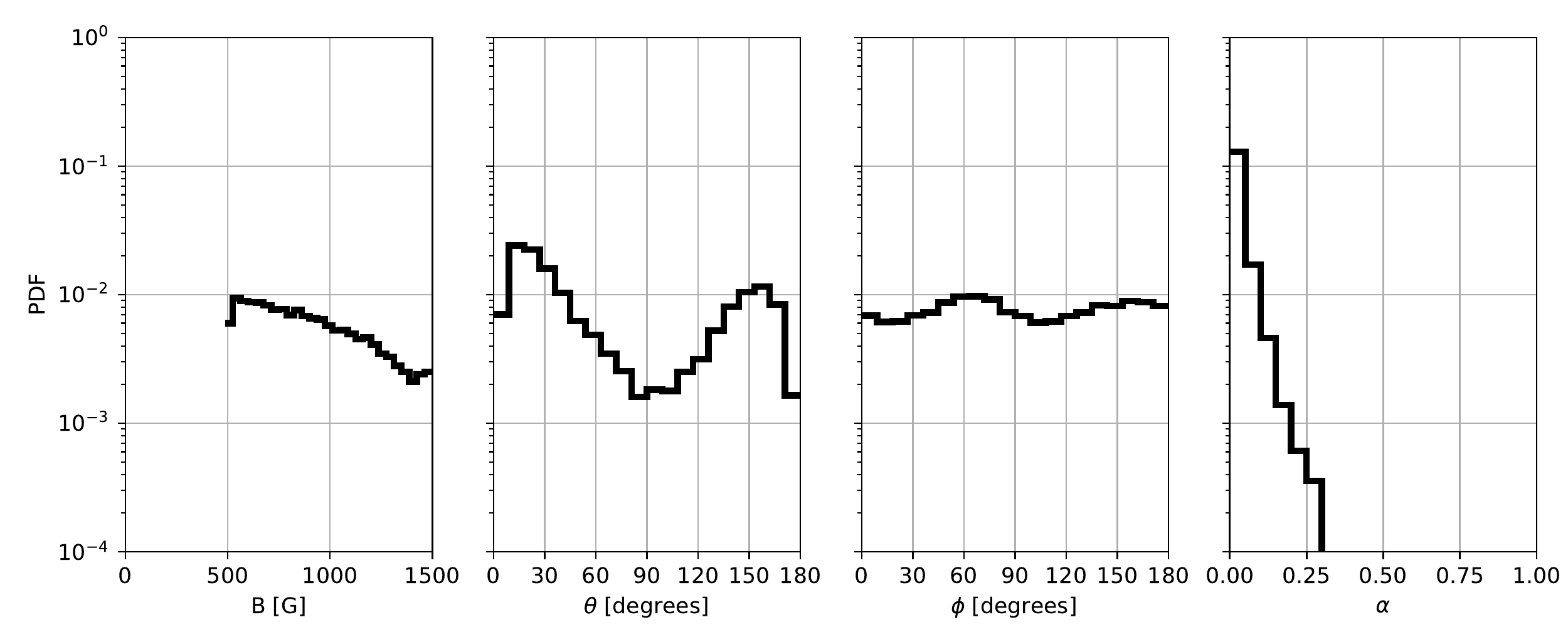}
     \caption{Statistical distributions of the magnetic parameters as inferred from the inversion of disc centre dataset. From left to right: Probability density functions for the LOS magnetic field strength (B), inclination ($\theta$), azimuth ($\phi$), and filling factor ($\alpha$). Here we present the pixels of the disc centre dataset that are classified as giving reliable information in the inversion process ($|B|>500$ G, see text).}
     \label{fig:magnstrenhistdc}
\end{figure*}

Figure \ref{fig:magnstrenhistdc} presents the magnetic field strength, LOS inclination, LOS azimuth, and filling factor probability density functions (PDF) for the disc centre. The magnetic field strength reaches from 500 G (our lower threshold) to about 1400 G. Inside this range, there is a marked preference for the lowest values in that range. These fields have an almost uniform PDF of the LOS azimuth, and a preference for the more vertical to the surface fields. The PDF of the filling factor is concentrated at low values (below 0.3), with a strong preference for the lowest
values ($\sim$0.02).

\begin{figure}
  \begin{center}
     \includegraphics[width=0.48\textwidth]{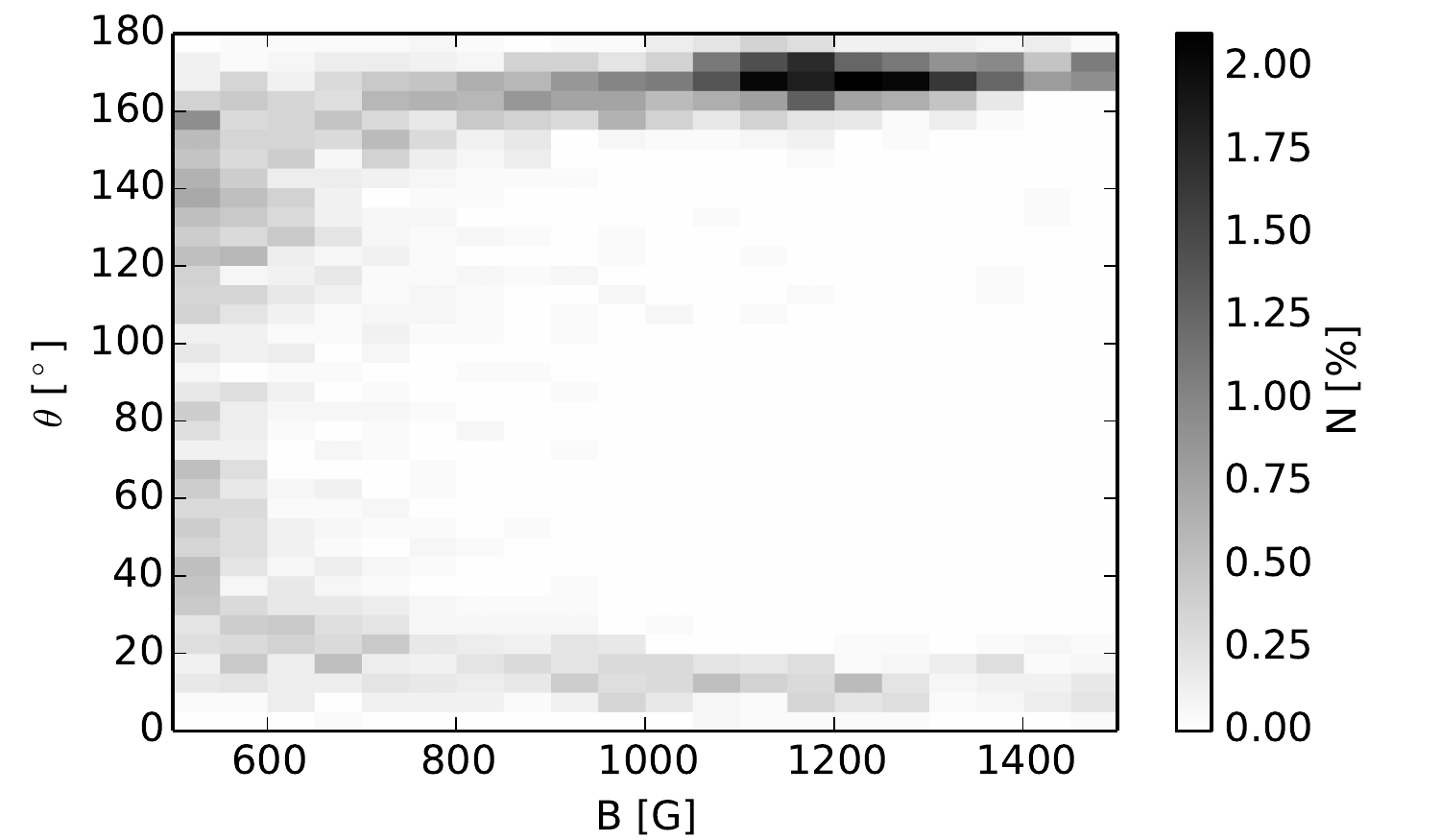}
     \caption{Bi-dimensional density plot of the magnetic field inclination as a function of the magnetic field strength. The bin sizes for the magnetic field strength and inclinations are 50 G and 5$^{\circ}$, respectively.
     }
     \label{fig:BvsINC_disccentre}
  \end{center}
\end{figure}

Figure \ref{fig:BvsINC_disccentre} shows the LOS magnetic field inclination dependence on their strength. The strongest fields are oriented close to the LOS direction, which at disc centre position is also vertical to the solar surface. This behaviour is well known to occur for the strongest QS magnetic fields \citep{steiner2007}. This vertical condition of the magnetic fields holds until 700 G, where the inclination distribution changes abruptly. Below this magnetic field strength value, the inclination distribution is evenly distributed between 10$^{\circ}$-80$^{\circ}$ and 100$^{\circ}$-170$^{\circ}$.
These inferred statistical properties are compatible with previous near-IR studies of the QS at disc centre \citep[see for instance][]{khomenko2003, martinezgonzalez2008}. 

\begin{figure*}
  \begin{center}
     \includegraphics[width=1.\textwidth]{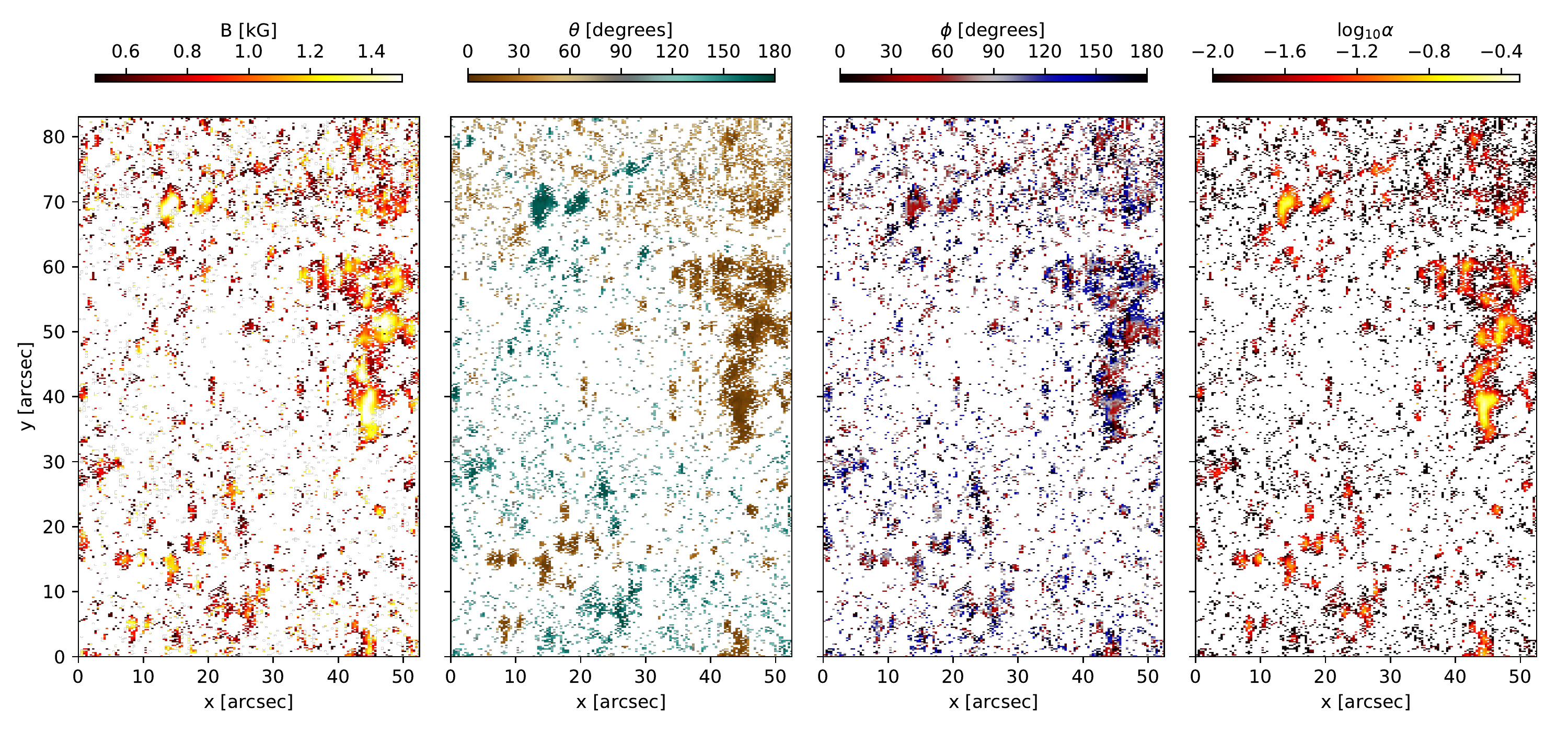}
     \caption{Disc centre spatial distribution for the reliably inferred LOS magnetic field parameters ($|B|>500$ G, see text). From left to right: LOS magnetic field strength (B), LOS magnetic field inclination ($\theta$), LOS magnetic field azimuth ($\phi$) and magnetic field filling factor ($\alpha$).}
     \label{fig:magnstren2Ddc}
  \end{center}
\end{figure*}

The spatial distribution of the same magnitudes throughout the FOV is presented in Fig. \ref{fig:magnstren2Ddc}. This figure
shows that the spatial distribution is not homogeneous, but forms patchy structures \citep{lites1996,martinezgonzalez2008b}. This applies for magnetic fields whose the topology is univocally inferred. For the whole polarimetric signals, the QS is permeated with magnetic fields in which no proper isolated magnetic patches can be distinguished. The strongest magnetic fields (e.g. those at x:48$^{\prime\prime}$, y:50$^{\prime\prime}$, or x:14$^{\prime\prime}$, y:70$^{\prime\prime}$) are associated with LOS inclinations closer to 180$^{\circ}$ (0$^{\circ}$), meaning that the strongest fields are the most vertical. Their filling factors are also the highest ($\sim0.4$), and the LOS azimuth seems to describe a radial expanding/converging configuration around them. There is wide agreement that these strong magnetic features are due to the vertical thin magnetic flux tubes \citep{spruit1976,schussler1990}. At the spatial resolution of these observations, it is not possible to spatially resolve their structure, but we can estimate the tube sizes using the inferred magnetic parameters. Our pixel size is $0.^{\!\!\prime\prime}35\times 0.^{\!\!\prime\prime}185$ and the mean filling factor value for these strong features is about 0.3. When we assume that the flux tubes have circular sections, these values give a flux tube radius of $\sim$117 km, which is in close agreement with the measured size observed with the highest spatial resolutions \citep{lagg2010} and with theory \citep{schussler1990}. The large (compared to hundreds of km) magnetic patches may be explained by two reasons. First, the observed structures are blurred by the point spread function of the seeing and of our optical system, which causes an unresolved structure to occupy several adjacent pixels. Second, these kG magnetic structures are usually observed in groups rather than individual structures. The joint action of these two effects may explain that we observe these structures as covering a large area of the observed FOV and that they form complex spatial shapes \citep[see, e.g.][]{requerey2015}. 

\begin{figure*}
\centering
     \includegraphics[width=1.\textwidth]{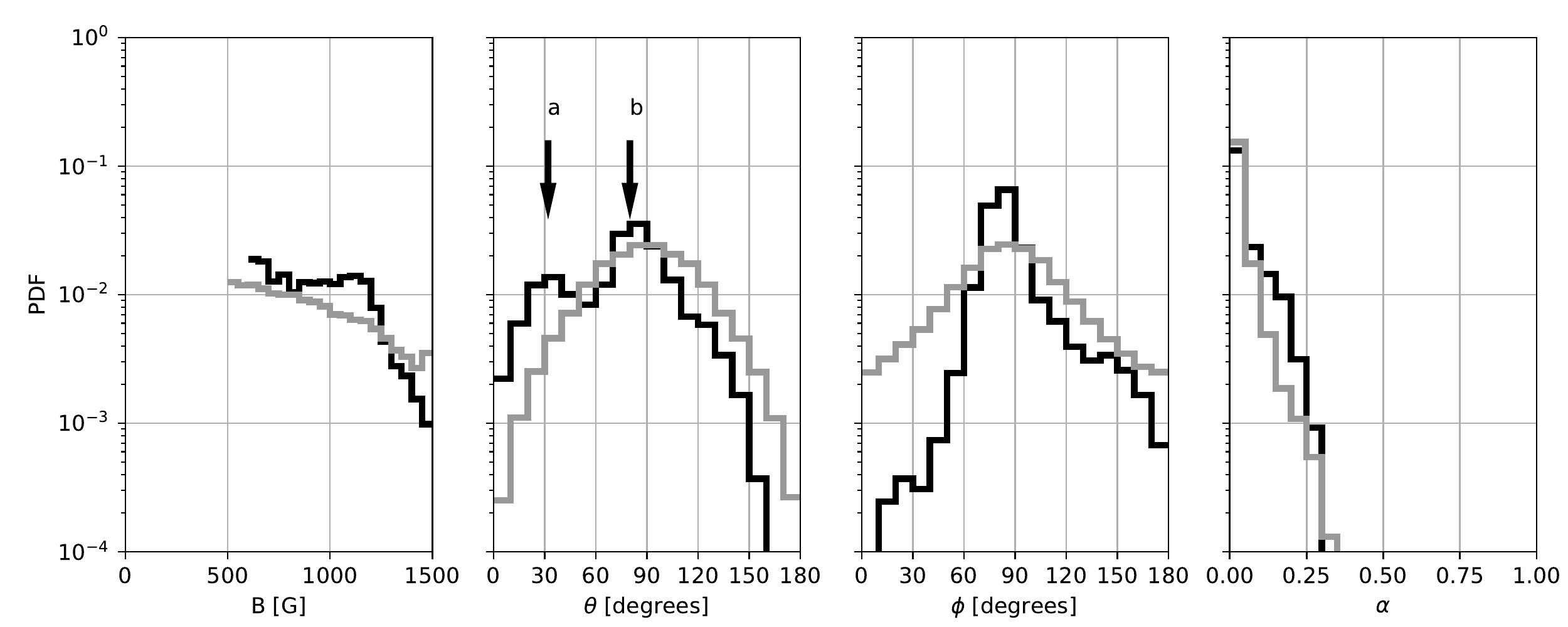}
     \caption{Statistical distribution of the inferred LOS magnetic field parameters at the west limb. Dark solid line: Probability density function of different magnetic parameters for the pixels that are considered to give reliable inferred parameters ($|B|>600$ G, see text). From left to right: Probability density functions for the magnetic field strength (B), inclination ($\theta$), azimuth ($\phi$), and filling factor ($\alpha$) in the LOS reference frame. In grey we show the same LOS magnetic field magnitudes for an LRF disc centre-like magnetic topology as would be observed at the west limb positions.}
     \label{fig:magnstrenhistwl}
\end{figure*}
\begin{figure*}
  \begin{center}
     \includegraphics[width=0.9\textwidth]{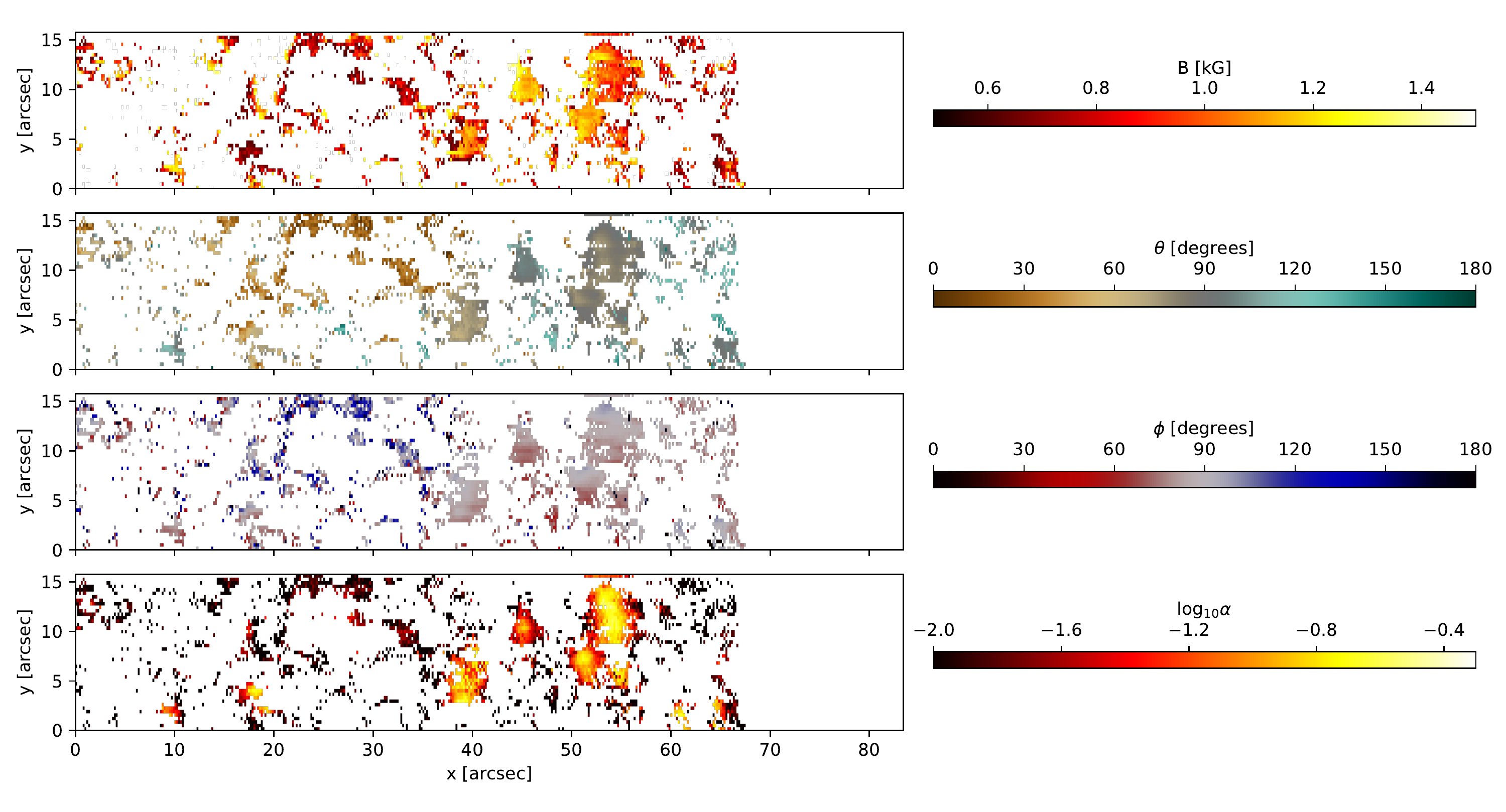}
     \caption{Same as Fig. \ref{fig:magnstren2Ddc} for the west limb dataset.
     }
     \label{fig:magnstrenwl2D}
  \end{center}
\end{figure*}

\subsection{West QS region}

Beyond the disc centre, the LOS direction and vertical directions no longer match. Hence, to interpret the LOS inclination and azimuth, we need to rotate the reference system from that of
the LOS to the local reference frame (LRF). The outcome of this change depends on the proper resolution of the $180^{\circ}$ LOS azimuth ambiguity. This change from LOS to LRF is described below. However, first, some insight into the magnetic topology of this QS region is required without such a strong dependence on the resolution of the LOS $180^{\circ}$ azimuth ambiguity.

In this step, we verified that the distributions of the magnetic field inclination and azimuth between the limb data and the disc centre were compatible. To this aim, we took the LOS magnetic field inclination and azimuth at the disc centre, which is also the LRF topology, and projected them to any disc position. We did not solve the LOS azimuth $180^{\circ}$ ambiguity for the
disc centre position, but randomly took one of the two possible LRF azimuths. This assumption is not strongly relevant in the calculation because for the disc centre, the inferred LOS azimuth distribution is homogeneous, so that from a statistical point of view, there is no reason to prefer one solution over the other. Moreover, since the field inclinations at disc centre are mostly vertical, the LOS azimuth $180^{\circ}$ ambiguity does not involve a critical change in the topology of the magnetic field in the LRF.

This method only considers limb effects that are due to the geometry of the viewing angle. We did not consider changes in the radiative transfer, the limb darkening, or different pixel size over the solar surface because they mainly affect the absolute amplitudes of the polarimetric signals, but not their relative amplitudes. We therefore expect these effects to be much smaller than those that are due to geometrical effects. Although these effects are not considered, this simple exercise gives some valuable insight into the comparison between the magnetic LOS topology at different disc positions.

Fig. \ref{fig:magnstrenhistwl} shows the observed west limb LOS magnetic field strength, inclination, azimuth, and filling factor PDFs (black). In grey (labelled ``synthetic'') we show the same parameter PDFs as they would be seen for a disc centre-like magnetic field distribution at that same solar disc position, taking into account the different viewing angle. We first focus on the change in disc centre-like magnetic field distribution. It is worth proceeding this way because we already know the topology of these magnetic fields at the LRF. The LOS inclination (second panel from the left of Fig. \ref{fig:magnstrenhistwl}) is strongly concentrated around 90$^{\circ}$, and the LOS azimuth is also concentrated around 90$^{\circ}$. In other words, the magnetic field vector is mostly contained in the plane of the sky (perpendicular to the LOS) and at 90$^{\circ}$ from the solar north-south direction (the reference for the azimuth). A radial magnetic field (such
as those at disc centre) will show these values of inclination and azimuth at the west limb. Therefore, the distribution of fields found in the disc centre, which were mostly vertical, correspond at the west limb to a distribution with a peak at 90$^\circ$ for the LOS inclination and azimuth. If this peak of inclinations is due to vertical fields, the presence of a dominant polarity involves a displacement of the peak from $90^{\circ}$. If the dominant polarity is positive (in the LRF), then the peak is shifted towards smaller LOS inclinations. If the dominant polarity is negative (in the LRF), then the peak moves to values
higher than $90^{\circ}$ .

\begin{figure*}
\centering
     \includegraphics[width=1.\textwidth]{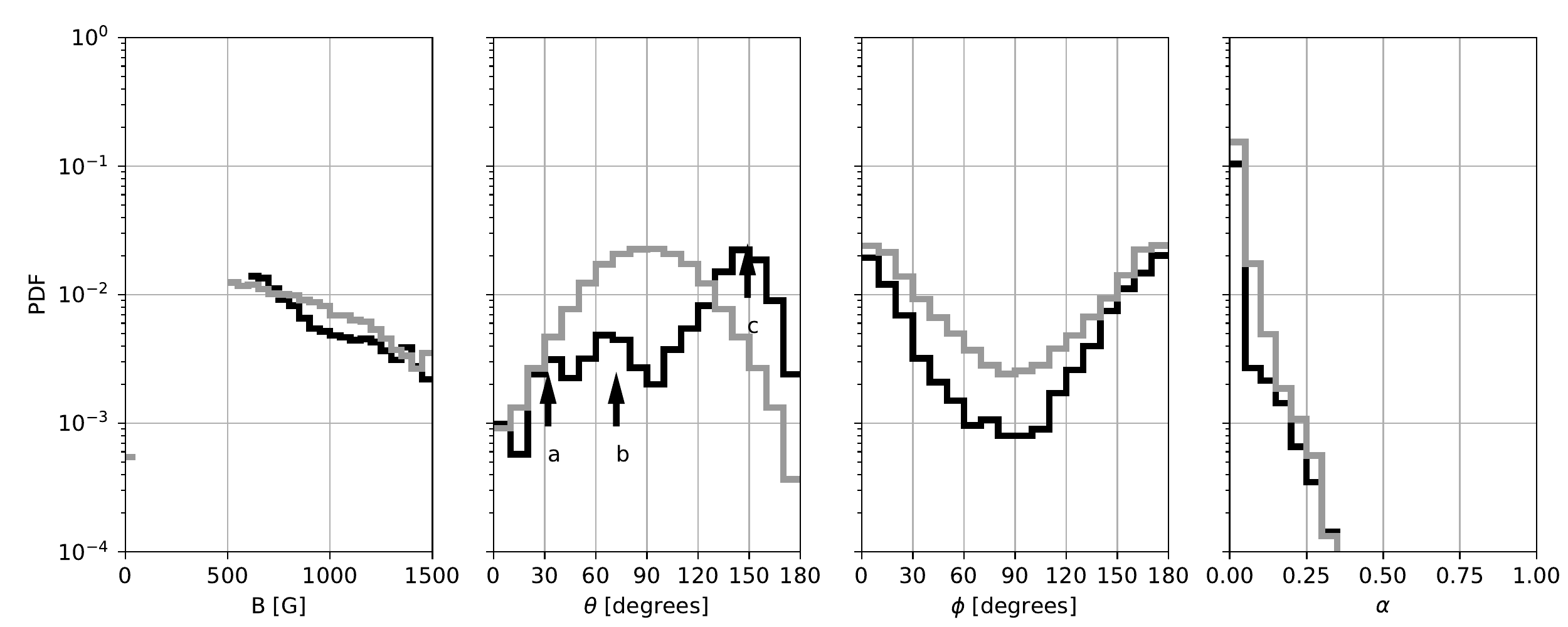}
     \caption{Same as Fig. \ref{fig:magnstrenhistwl} for the north region dataset.}
     \label{fig:magnstrenhistnp}
\end{figure*}

\begin{figure*}
  \begin{center}
     \includegraphics[width=1.\textwidth]{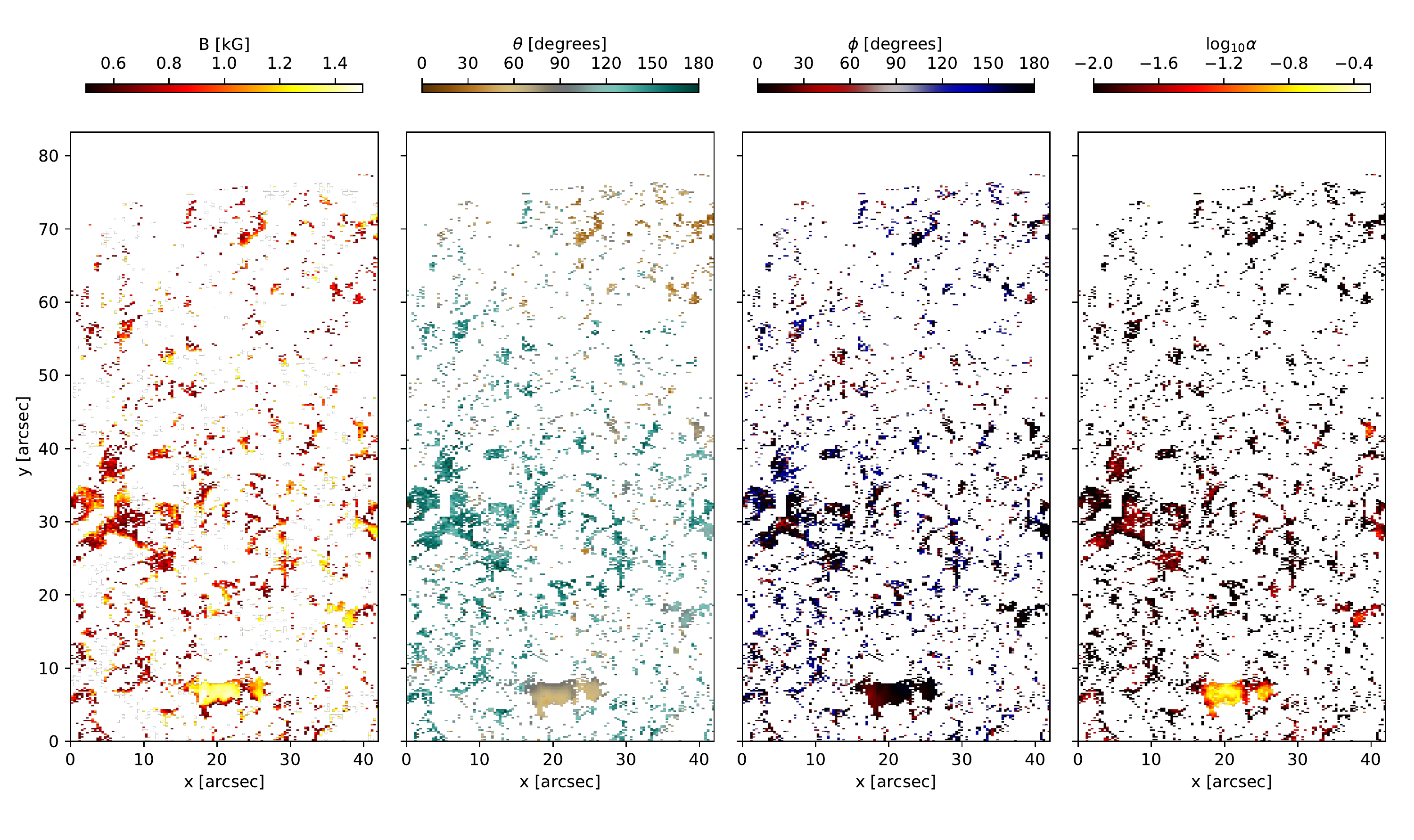}
     \caption{Same as Fig. \ref{fig:magnstren2Ddc} for the north region dataset.}
     \label{fig:magnstrennp2D}
  \end{center}
\end{figure*}

The LOS magnetic field strength, azimuth, and filling factor distributions show, in general terms, a good agreement between the observed parameters for the west limb (black lines), and what would be seen for a disc centre-like topology at that same disc position (grey lines). For the west limb data (black line),
the LOS inclinations peak close to 90$^{\circ}$, and the azimuths are also concentrated around $90^\circ$. These fields are compatible with the presence of vertical magnetic fields. The west LOS inclination peak is slightly shifted to lower values ${\sim}80^{\circ}$. Another interesting feature observed at the west limb is a secondary peak (see arrow {\it a} in Fig. \ref{fig:magnstrenhistwl}) in the LOS inclination distribution at around 30$^{\circ}$, which has no observed counterpart at the disc centre, as deduced from the shape of the LOS inclination for a disc centre-like magnetic topology. The reason might be that we observe a magnetic component that is absent from disc centre or is due to the observational effects close to the limb over observed magnetic features at disc centre. This point is further considered in Sec. \ref{ssection:noncompatible}.

Figure \ref{fig:magnstrenwl2D} shows the spatial distribution of the magnetic parameters. As for the disc centre, magnetic fields appear in patchy structures scattered throughout the FOV. The strongest magnetic field strengths are associated with the highest filling factors, in a similar way as for the disc centre. At this disc position, the strongest magnetic patches exhibit a decreasing magnetic field strength from the side of the structure facing disc centre to limbward (see  x:54$^{\circ}$, y:11$^{\circ}$ for a very clear example). Furthermore, the magnetic filling factor in these structures is characterised by lower values at both sides, the one facing disc centre, and the one limbward. These magnetic features show an LOS azimuth that gradually moves from lower than 90$^{\circ}$ at the bottom part of the structures (redish colour) to values slightly above 90$^{\circ}$ in the upper part of the structure (bluish colour).

\subsection{North QS region}

As for the west limb data, in the north region, the structures
vertical to the surface and the LOS are not aligned. For this reason, we repeated the same experiment for this solar disc position (see the previous section). The result is presented in Fig. \ref{fig:magnstrenhistnp} in grey, together with the observed distributions for the north data. For the ``synthetic'' distributions, the LOS inclinations behave very similar as at the west limb, namely, an LRF vertical magnetic field distribution gives rise to a prominent LOS inclination of 90$^{\circ}$ close to the limb. In contrast, the LOS azimuth retrieved now peaks at 0/180$^{\circ}$. This is again compatible with vertical fields since the reference of the azimuth is the north-south solar direction, hence aligned with the radial direction to the limb. In the more general case, the LOS azimuth of a vertical magnetic field at the very limb is related to the angle between the radial direction and the azimuth origin.

As in the case of the west region, magnetic field strength, LOS azimuth, and filling factor again show a behaviour similar to what is expected for a disc centre-like magnetic topology. As in the west case, these are also some differences in the LOS inclination distribution. In particular, instead of the smooth distribution of a disc centre-like case, the north region LOS inclination distribution is characterised by three localised bumps. The one highlighted with {\it b} resembles what is expected for vertical fields in the LRF. The other two peaks, marked  {\it a} and {\it c} in Fig. \ref{fig:magnstrenhistnp}, share properties between them and with the previously seen fields labelled {\it a}  for the west region dataset.

The spatial distribution of the magnetic parameters of the north QS region dataset are presented in Fig. \ref{fig:magnstrennp2D}. Some magnetic patches are consistent with the topology expected
from a disc centre-like one. For instance, the strongest magnetic fields are characterised by the highest magnetic filling factors. These fields are characterised by LOS inclinations close to $\sim90^{\circ}$ ({\it b} in Fig. \ref{fig:magnstrenhistnp}), and their LOS azimuths are mostly along the N-S direction. In contrast to the west region, most of the magnetic patches have LOS magnetic field inclinations of about 30$^{\circ}$ and 150$^{\circ}$ ({\it a} and {\it c} in Fig. \ref{fig:magnstrenhistnp}). 

We further explored this bi-modal behaviour seen at the LOS magnetic field topology by rotating from the LOS to the LRF systems. To exemplify this procedure and emphasise the differences between the two modes, we considered two magnetic structures that resemble each behaviour.

\begin{figure*}
   \centering
     \includegraphics[width=1.\textwidth]{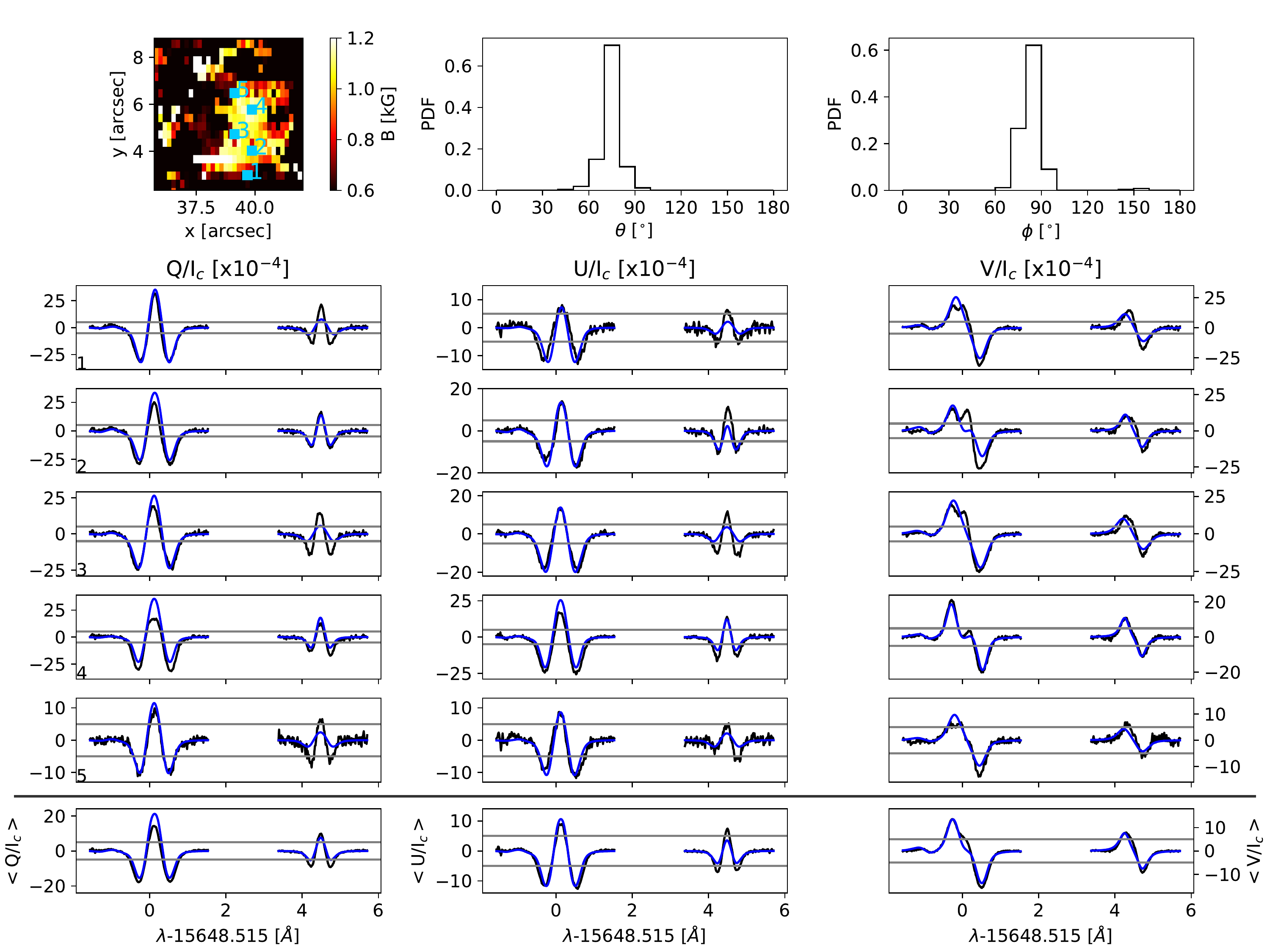}
     \caption{Different parameters found for a kG magnetic field structure observed in the west dataset. Top row, from left to right: Magnetic field intensity map, the probability density function of the LOS magnetic field inclination, and the LOS magnetic field azimuth probability density function. The panels below, from left to right, show the Stokes Q, U, and V observed profiles (dark line) and inversions (blue lines) for the various pixels highlighted in the image. The bottom row shows the observed (black) and inverted (blue) profiles averaged over the whole structure. The grey horizontal lines represent the $\pm5\sigma$ levels.}
     \label{fig:wlstronglineal}
\end{figure*}

\subsection{Magnetic fields compatible with vertical fields}

   \begin{figure*}
   \sidecaption
   \includegraphics[width=0.7\textwidth]{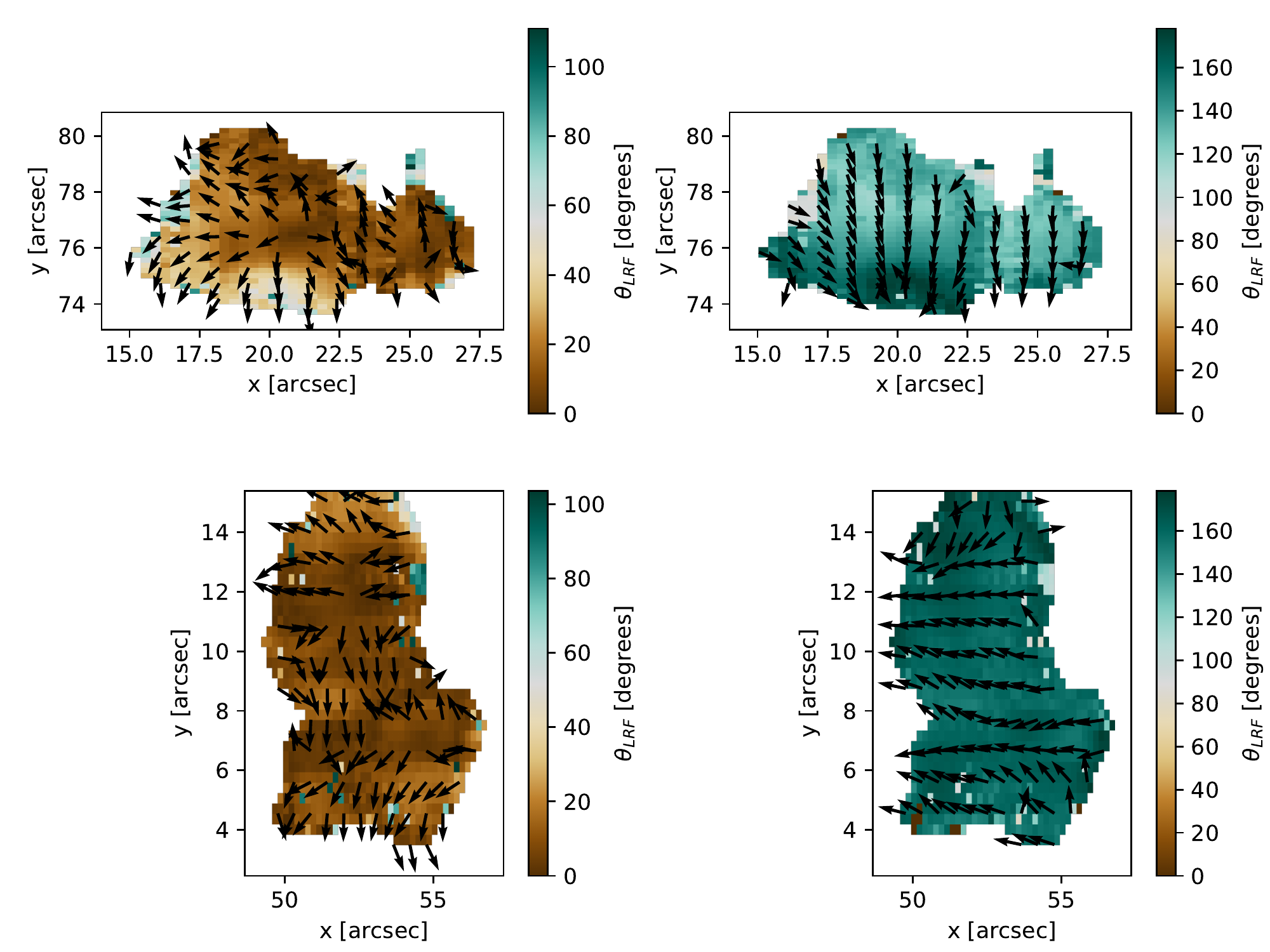}
      \caption{LRF magnetic topology of two magnetic patches extracted from limb datasets. From top to bottom: Two structures taken from the north region dataset and from the west region, respectively. The magnetic field inclination is colour-scaled. Arrows show the magnetic field azimuth. Arrows are plotted for one of every three pixels in each direction for clarity. Each column represents one of the two possible solutions retrieved due to the LOS magnetic field azimuth 180$^{\circ}$ ambiguity.}
         \label{fig:locmagneticstructures}
   \end{figure*}

An example of a typical structure characterised by the peak labelled {\it b} in Fig. \ref{fig:magnstrenhistwl} and \ref{fig:magnstrenhistnp} is presented in Fig. \ref{fig:wlstronglineal}. This structure belongs to the west region dataset, but similar examples can be found in the north region. The magnetic field strength spatial distribution of the structure is displayed in the top left panel. It covers a projected area of $\sim2^{\prime\prime}\times4^{\prime\prime}$ ($\sim9^{\prime\prime}\times4^{\prime\prime}$, when correcting for foreshortening) and is characterised by magnetic field strengths ranging from 800 G to 1.2 kG. The Stokes $Q$, $U,$ and $V$ spectra for the highlighted pixels are shown in the lower panels of Fig. \ref{fig:wlstronglineal}, as well as the average Stokes profiles over the structure (bottom row of panels). These profiles show strong linear polarimetric signals (horizontal grey lines represent the noise threshold of $5\sigma$ ) with strong Stokes $V$ signals
as well. The linear polarisation profiles reach higher amplitudes than those of the Stokes $V$ profiles. This picture is hardly visible at the disc centre and is due to the perspective. The presence of vertical fields, which are widely observed at disc centre, can explain these fields \citep[see, e.g.][and references therein]{solanki2009}. The LOS magnetic field inclination distribution (top middle panel), which is clearly dominated by values of about 75$^{\circ}$ , and the LOS magnetic field azimuth (top right panel),
which prefers values around 90$^{\circ}$ , are also expected for vertical magnetic fields. To confirm that these magnetic fields are vertical, we proceeded to rotate the LOS to the LRF.

Before rotating from LOS to LRF, we point out that Stokes V profiles with more than two lobes are frequent. This feature is characteristic of solar atmospheres more complex than atmospheres where an unique and constant in LOS magnetic field element is present. An example of this feature is shown in the second profile of Fig. \ref{fig:wlstronglineal}, where the Stokes V profile shows a third lobe. We checked that when we increase the complexity of the model atmosphere (in particular, using two magnetic components constant in LOS), the additional lobe/s are fitted. To do so, the magnetic components inferred are clearly separated into two groups: a strong (kG) component, and a weak (hG) one. The comparison of the topology of this strong component (the weak one is below our threshold), with the component obtained for a single magnetic element shows not significant change in the topology of the magnetic field because the linear profiles of these cases are mostly due to the strong field ,while Stokes V is a mixture of both components. Hence, when using only one magnetic component, the magnetic properties are inferred mostly from the Stokes linear profiles and part of Stokes V profiles, which strongly restricts the effect of the ``weak'' component on the latter.

The rotation from the LOS to the LRF systems can be achieved with a single rotation of the heliocentric angle to the position of the pixel considered. Instead, it is worth proceeding with two rotations in order to ensure that the azimuth reference is kept at the north solar direction and increasing eastwards. The most delicate problem is solving the 180$^{\circ}$ ambiguity of the LOS azimuth, which implies two possible configurations of the magnetic field in the LRF for every pixel. Several methods have been developed to solve this problem \citep[e.g. ][]{metcalf2006}. Here we have followed an alternative method. In a first step, we rotated from the LOS to the LRF, keeping the two possible solutions due to the LOS azimuth ambiguity:

\begin{itemize}
  \item calculate the helioprojected solar longitude and latitude of each point,
  \item derive the two possible LRF configurations for each pixel, associated to the two possible values of the LOS azimuth,
  \item identify magnetic patches whose structure is assumed to be coherent according to the field strength patching,
  \item for each structure, choose a starting point randomly with a given solution of the two possible points,
  \item for each structure, recover the general LRF solution that minimises the difference between the LRF magnetic field with the adjacent LRF configurations inside the structure. Given a pixel, the difference is calculated as the sum of the magnetic vector modulus difference with its adjacent pixels.
\end{itemize}

Following this analysis, we analysed eight (seven) structures that cover 4.52\% (12.21\%) of the north (west) limb dataset. This yielded two possible magnetic configurations for each magnetic patch (see, e.g. Fig. \ref{fig:locmagneticstructures}), depending on the solution adopted for the starting point.

The second step is to choose one of the retrieved solutions. To do so, we chose as the more likely structure the structure that contributes the final LRF azimuth (given by all detected structures) as homogeneously as possible. This was done by determining the most homogeneous final LRF azimuth distribution.

   \begin{figure*}
   \centering
       \includegraphics[width=1.\textwidth]{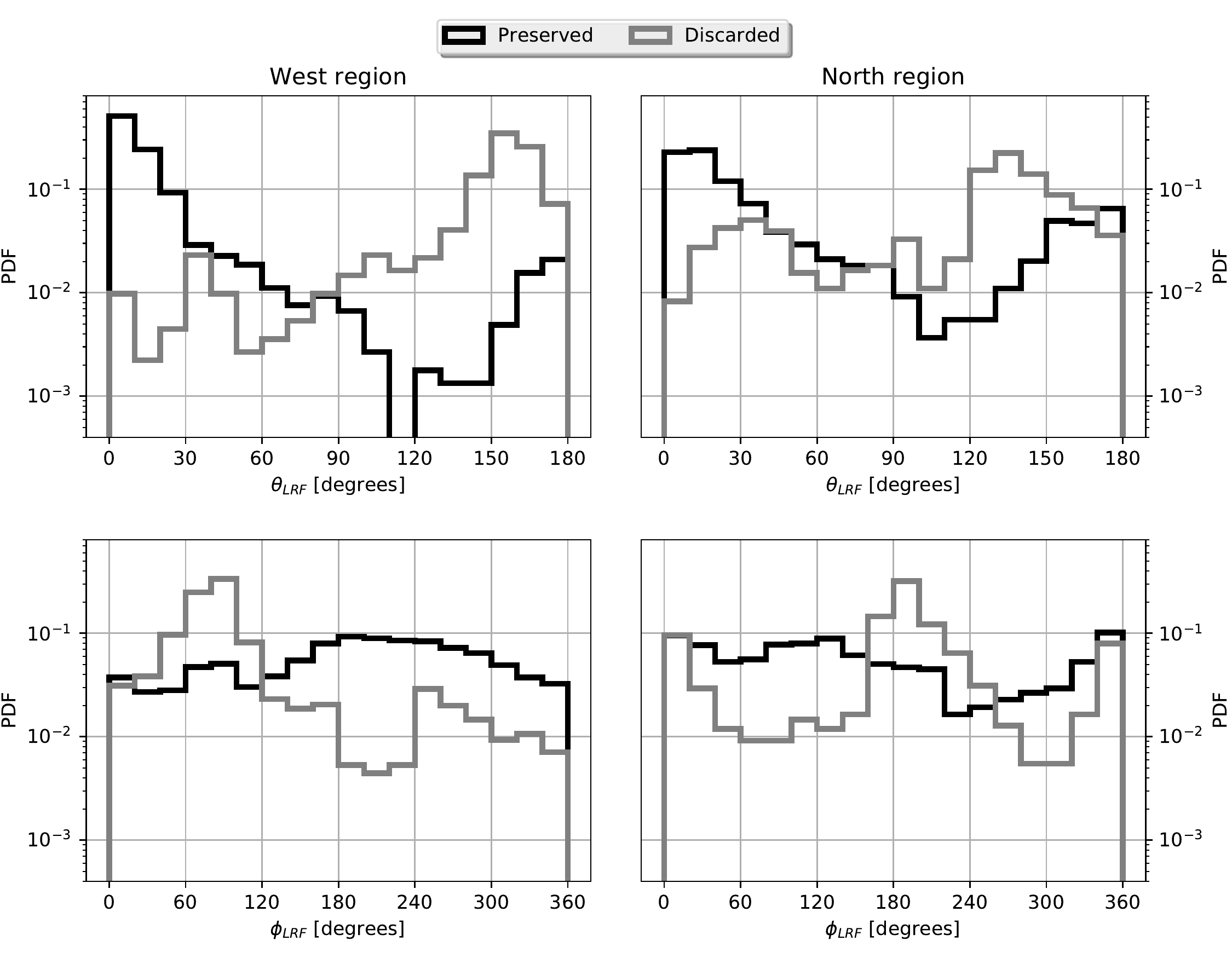}
       \caption{Statistical distributions of the LRF magnetic field at the limb data. Left panels, from top to bottom: Probability density functions of the preferred magnetic field inclination ($\theta_{LRF}$) and azimuth ($\phi_{LRF}$) for the north region dataset. Right panels, from top to bottom: Probability density functions of the discarded magnetic field inclination ($\theta_{LRF}$) and azimuth ($\phi_{LRF}$).}
       \label{fig:locNPmagneticdistributions}
   \end{figure*}

\paragraph{Preferred distribution.} The preserved LRF inclination and azimuth distributions are presented in Fig. \ref{fig:locNPmagneticdistributions} in solid black. For this LRF inclination distribution, the magnetic fields are characterised by vertical magnetic fields. In both the north and west dataset lie magnetic fields of both polarities with a dominant positive polarity. This is consistent with the above-mentioned argument about the shift of the LOS inclination distributions to values lower than 90$^{\circ}$ . With the data we analysed, we cannot assess if this dominant polarity of the dataset is representative of the remaining PR or if it is a statistical fluctuation due to the small FOV observed. We recall that we detected almost ten magnetic structures of this type of magnetic fields at each limb FOV. The retrieved LRF azimuth for these vertical fields is quite flat. This topology resembles that of the strongest magnetic fields at disc centre well. This means
that a coherent scenario exists that can explain the inferred magnetic topology for this subset of magnetic fields at the three disc positions.

\paragraph{Discarded solution.} The discarded solution is shown as the grey solid line in Fig. \ref{fig:locNPmagneticdistributions}. The LRF azimuth distribution is strongly aligned with the radial solar direction. At the west limb, this means that the LRF azimuth peaks at 90$^{\circ}$ (most of them) and 270$^{\circ}$. At the north limb, this direction is given by 0$^{\circ}$ and 180$^{\circ}$ LRF azimuth values. The LRF inclination distributions we discarded are characterised by a strong preference for an LRF inclination of 30$^{\circ}$ and 150$^{\circ}$ . This discarded topology has a preferred orientation aligned with the solar radial direction and at mid-inclined fields. Since this is a highly singular orientation for which we find no reason, the preferred solution is probably
more likely.

\subsection{Magnetic fields that are incompatible with vertical fields}
\label{ssection:noncompatible}

\begin{figure*}
   \centering
     \includegraphics[width=1.\textwidth]{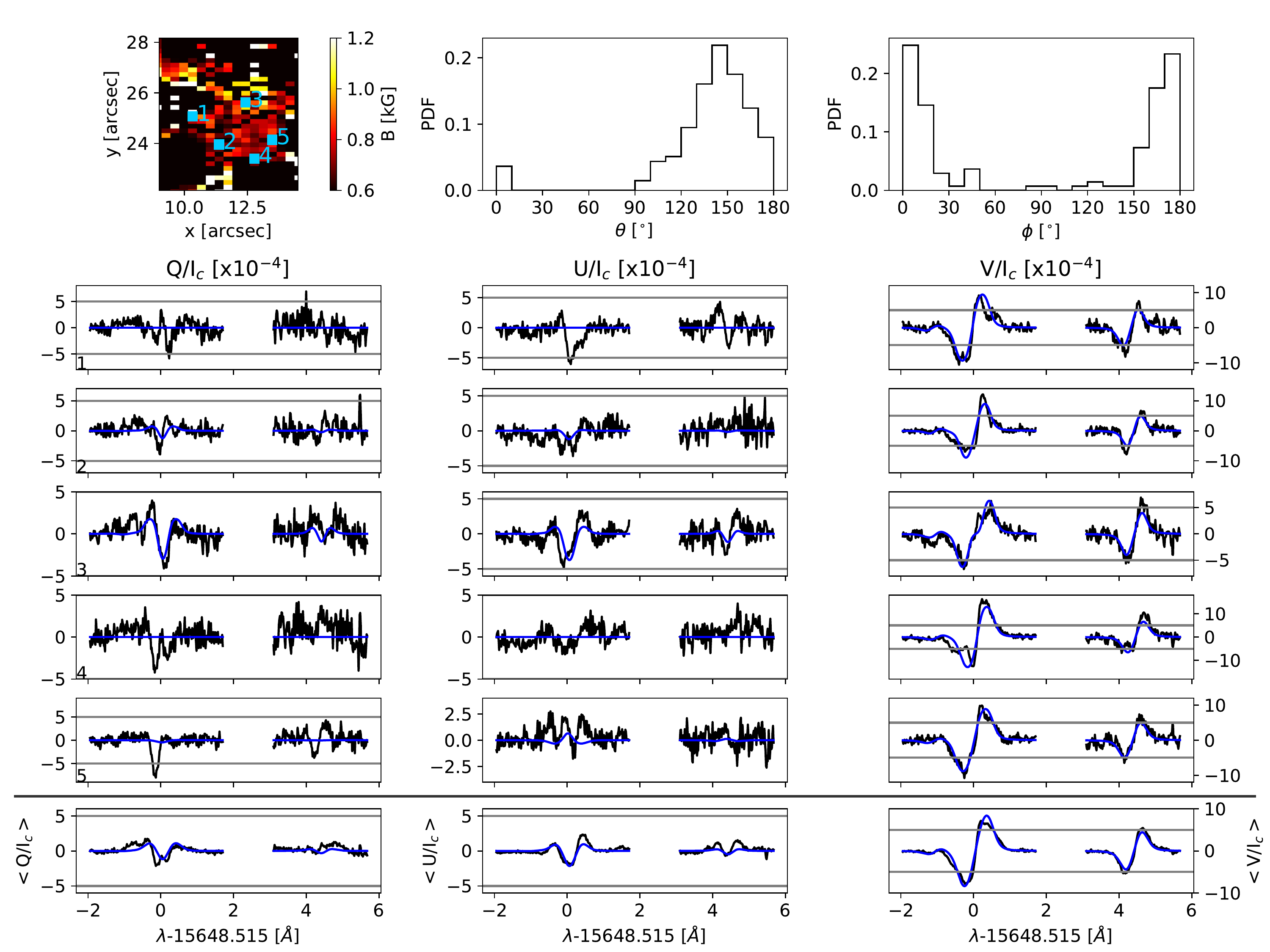}
     \caption{Same as Fig. \ref{fig:wlstronglineal}, but for a different structure in the north limb dataset.
     }
     \label{fig:npweaklineal}
\end{figure*}

Consider now an example of a magnetic structure whose topology is representative of mode {\it a} in Fig. \ref{fig:magnstrenhistwl} and of {\it a} and/or {\it c} in Fig. \ref{fig:magnstrenhistnp}. This is shown in Fig. \ref{fig:npweaklineal}. This magnetic patch is similar in size to that of Fig. \ref{fig:wlstronglineal}, covering a projected area $\sim3^{\prime\prime}\times2^{\prime\prime}$ ($\sim3^{\prime\prime}\times11^{\prime\prime}$ when corrected
for projection effects). This structure also shows Stokes $V$ polarisation signals above our $5\sigma$ threshold, reaching amplitudes of the order of $1\times10^{-3}$. However, it clearly differs from the structure in Fig. \ref{fig:wlstronglineal}: the linear polarisation signals are barely above the noise level. Despite the weak signals, the inferred azimuth for the structure is highly aligned in the E-W direction (top right panel in the figure). This situation is not expected if Stokes $Q$ and $U$ are pure noise, so this might be indicative of a coherent, although very weak, information content in the linear polarisation profiles. If this indeed is the case, the average of the linear polarisation profiles within the structure should present a Zeeman-like signature above the noise level. This average is presented in the last row of Stokes profiles in Fig. \ref{fig:npweaklineal}. The Stokes $Q$ and $U$ polarisation profiles both have familiar Zeeman shapes and the averaged inversion fits them quite well, meaning that the code is able to catch these features despite the faintness of the signals. It is clear that these fields cannot be explained by a purely vertical structure as can the fields observed at disc centre, since vertical fields close to the limb involve linear polarisation signals of the order of or stronger than the circular ones.

To explain these structures, we first considered that such a structure is due to a single magnetic field orientation, that
is, that the observed topology for each pixel can be explained by a single magnetic field configuration. For the sake of simplicity, we consider as a particular example a magnetic field with an LOS inclination of 150$^{\circ}$ and an LOS azimuth of 0$^{\circ}$, observed at a helioprojected latitude of 75$^{\circ}$ and a helioprojected longitude of 0$^{\circ}$. These values match the average parameters of the structure shown in Fig. \ref{fig:npweaklineal}. In this case, the two possible LRF solutions for such a magnetic field configuration are 75$^{\circ}$ and 135$^{\circ}$, depending on the two possible LOS azimuths due to the 180$^{\circ}$ ambiguity. One of the two possible solutions is strongly horizontal ($\theta_{LRF}\sim75^{\circ}$). This solution is a highly unexpected configuration for two main reasons: 1) According to theory, strong magnetic fields (the observed fields range from 600 G up to 1000 G) would tend to vertical configurations \citep{spruit1976,schussler1990}. 2) the QS magnetism observed at the disc centre is dominated by vertical fields for these magnetic field strengths (at disc centre, only magnetic fields below 700G are not exclusively vertical). The other possible local magnetic field configuration for the averaged values stated above ($\theta_{LRF}\sim135^{\circ}$) is given by magnetic fields inclined ${\sim}$45$^{\circ}$ and aligned with the radial direction, that is, the line joining the position where the magnetic structure is observed and the centre of the solar disc. This second magnetic field configuration describes a very particular orientation in the two limb datasets. This type\ of magnetic field topology is not observed at disc centre. These two facts seem to indicate that some projection effect plays a significant role in the observed profiles and in the subsequent inferred parameters. In particular, the spatial
resolution is poorer the closer to the limb it is, which might help understand this magnetic field topology. At low spatial resolution at large heliocentric angles, we might be mixing different magnetic fields in the same pixel.

\begin{figure*}
   \centering
     \includegraphics[width=1.\textwidth]{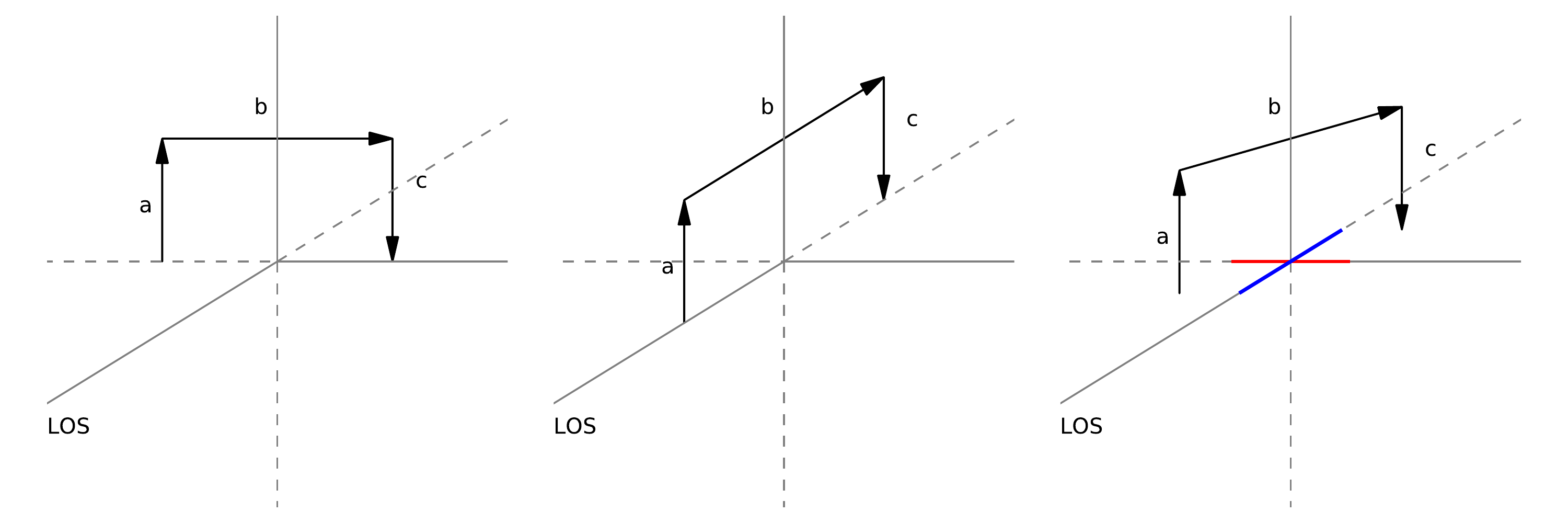}
     \caption{Schematic drawing of three different configurations of a staple-like loop at the very limb of the Sun. From left to right: Loop perpendicular to the LOS, loop aligned with the LOS, and loop at 45$^{\circ}$ from the LOS. In the latter, red line is the projection of b to the perpendicular to the LOS direction and blue line is the projection of b along the LOS direction.
     }
     \label{fig:loops}
\end{figure*}

In the past years, several reports of observations of small-scale magnetic loops in the QS have been published \citep{martinezgonzalez2007, centeno2007, martinezgonzalez2009, ishikawa2010, martinezgonzalez2010, gomory2010, wiegelmann2010, martinezgonzalez2011, mansosainz2011, martinezgonzalez2012, guglielmino2012, viticchie2012, gomory2013}. These structures consist of two footpoints of opposite polarity that are linked by an arc-like magnetic system whose size varies from sub-arcseconds to a few arcseconds. Magnetic loops appear in various shapes and magnetic field strengths, and their emergence is not homogeneous in time or space \citep{martinezgonzalez2012}. When observing at the limbs, the foreshortening strongly reduces our spatial resolution and these structures are likely unresolved, which poses the question whether an unresolved magnetic loop might explain an observation such as in Fig. \ref{fig:npweaklineal}.

We first considered a very simple situation to gain qualitative insight into this scenario. Consider a loop given by three magnetic vectors, {\it a}, {\it b,} and {\it c} (see left panel in Fig. \ref{fig:loops}). Two vectors are vertical and of opposite polarity ({\it a} and {\it c}), labelled as footpoints. The third vector ({\it b}), labelled as the apex, would be horizontally connecting both footpoints. For the sake of simplicity, we considered this loop to lie at the very limb with three different configurations:

\begin{itemize}
\item The loop is perpendicular to the LOS (left panel in Fig. \ref{fig:loops}). In this case, no magnetic field vector would give rise to circular polarisation, as the three vectors are perpendicular to the LOS. Both footpoint magnetic vectors ({\it a} and {\it c}) give rise to linear polarisation signals in the radial direction. If we set the origin of magnetic azimuth in positive Stokes $Q$ and aligned with the radial direction, then both footpoints give rise to positive Stokes $Q$ signals. On the other hand, the {\it b} magnetic field would be perpendicular to the previous case, and hence would give rise to negative Stokes $Q$ signals. For the particular case of this configuration, an unresolved loop would therefore tend to show no polarimetric signal.
\item A second scenario is obtained when the loop is aligned with the LOS (middle panel in Fig. \ref{fig:loops}). In this case, both footpoints ({\it a} and {\it c}) again would give rise to positive $Q$ signals as they are aligned with the radial direction. Now, the apex vector ({\it b}) is parallel to the LOS direction, however, and gives rise to Stokes $V$ polarisation signals (the sign depends on whether the loop points towards us, positive Stokes $V,$ or not, negative Stokes $V$). In this case, the unresolved loop would give rise to both Stokes $V$ and $Q$, with their amplitudes depending on the heights that the spectral line is sensitive to and to the particular configuration of the loop.
\item The last scenario is an intermediate case, in which the loop has an orientation between the previous two (right panel in Fig. \ref{fig:loops}). Now, both footpoints ({\it a} and {\it c})  again yield linear polarisation signals in the radial direction (positive Stokes $Q$). However, in this intermediate orientation, the apex vector ({\it b}) has magnetic components projected to both the LOS and the perpendicular to it, that is, this time, the apex gives circular and linear polarisation at the same time. The LOS component (blue in the figure) gives Stokes $V$ polarisation signals. The component perpendicular to the LOS (red in the figure) gives rise to negative Stokes $Q$, as it is perpendicular to the radial direction. In this way, the intermediate scenario gives Stokes $V$ signals (the apex) and Stokes $Q$ signals that are given by the contribution of both footpoints minus the contribution of the linear polarisation signal of the apex, hence they are typically very weak.
\end{itemize}

\begin{figure*}
   \centering
     \includegraphics[width=1.\textwidth]{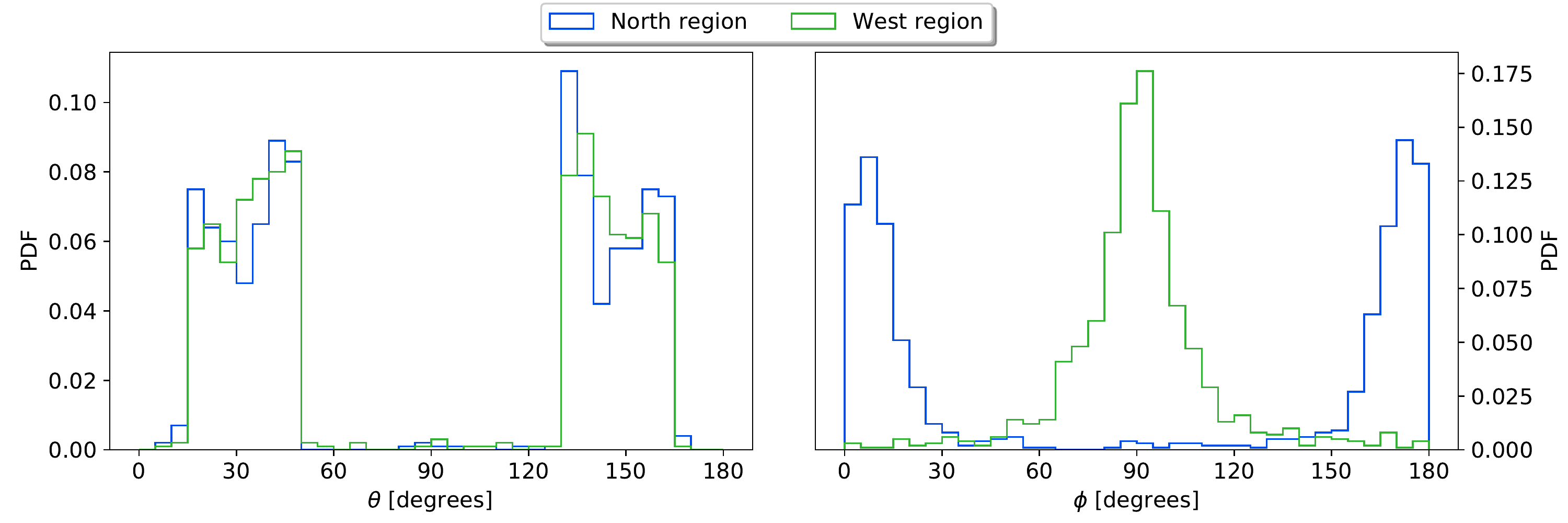}
     \caption{Simulated LOS inclination and azimuth distributions for magentic loops distributed at the two limb observations. Left: Probability density functions of the LOS magnetic field inclination inferred for 1000 synthetic unresolved magnetic field loops in the north (blue) and in the west region (green). In the right panel, we show the same representation for the LOS azimuth in between 0$^{\circ}$ and 180$^{\circ}$. The azimuth origin is in the solar north-south direction.}
     \label{fig:numericalloops}
\end{figure*}

We recall the VTT polarimetric calibration takes the terrestrial N-S direction as the reference for positive Stokes $Q$. Therefore we observe Stokes $Q$ and $U$ signals for the solar radial direction. The change from Earth to the solar reference frame requires taking account the P0 angle, which for the observation date was 23.$^{\!\!\circ}$88.

This staple-like loop is convenient to have as a readily available interpretation of the contribution from each part of the loop. However, more realistic loop shapes (like those in \citealt{martinezgonzalez2010} or \citealt{ishikawa2010}) have to be considered. Furthermore, the above qualitative argument is valid at the very limb, but our data cover solar surface areas whose heliocentric angles reach 70$^{\circ}$. In both cases, we expect that more realistic models would change the relative weights of the contributions from the various loop parts, but change them not enough to invalidate the general picture we considered. In order to support this statement, we present the following numerical test.

For this test, we considered a loop given by a semi-circle of constant magnetic field strength. The synthetic magnetic loops are numerically placed at the north and west limbs. The orientation of the magnetic loops, given by the line joining both footpoints with respect to the radial direction, is taken randomly in any direction. For each magnetic loop, we transformed the magnetic field vector from the LRF to the LOS. We synthesised the complete Stokes profiles for each of the points belonging to the loop for an HSRA atmosphere. We took the average of the Stokes profiles and inverted them to infer the magnetic field vector for the unresolved loop. This inversion was performed with the magnetic field strength, LOS inclination, LOS azimuth, and filling factor as free parameters, fixing all the thermodynamical parameters and ascribing the origin of LOS azimuths at the N-S solar direction. We performed this experiment for 1000 magnetic loops in the north solar region and for another 1000 loops in the west region. For each of these loops the field strength is taken to be constant and in between 600 G and 1000 G. This range is chosen from the typical magnetic field strength values that are characteristic of the observed magnetic structures. 

Figure \ref{fig:numericalloops} shows the distributions retrieved from the inversion for the LOS magnetic field inclination (left) and azimuth (right). In the two solar disc regions, the LOS inclination is located at specific values: around 30$^{\circ}$ and 150$^{\circ
}$ , and the LOS azimuth distribution is strongly aligned with the radial direction. This orientation is given by LOS azimuths of 0$^{\circ}$ and 180$^{\circ}$ in the north region (blue in the Fig \ref{fig:numericalloops}) and by 90$^{\circ}$ in the west region (green in the Fig \ref{fig:numericalloops}). Qualitatively, this model also explains the observed polarimetric signals. The averaged polarimetric profiles for each loop are characterised by weak linear magnetic signals together with clear Stokes V spectra.

This unresolved magnetic loop scenario and the simplified scenario explained above are able to reproduce the observed features of structures like the one in Fig. \ref{fig:npweaklineal}: Firstly, the circular polarisation signals; and secondly, the linear polarisation signals, given by the competition of the linear signals from the footpoints and apex. This competition might give rise to weak signals that contain enough information of the azimuth direction of the magnetic field. This model also explains the preferred radial orientation of the inferred LOS azimuths, and also the $\sim$30$^{\circ}$ and $\sim$150$^{\circ}$ LOS inclination preference.

\section{Discussion and conclusions}
\label{Sect:discussion_conclusions}

We have carried out an analysis of full spectropolarimetric observations of three different QS regions on the solar disc. The analysed spectral lines are two highly sensitive magnetic Fe\,{\sc i} lines located close to 1.565 $\mu$m. These lines are excellent candidates for magnetic field studies as magnetic sensitivity increases with wavelength and their Land\'e factors are high. The deep-mode observations have allowed us to detect magnetic signals in more than 70\% of each field of view. This is in agreement with recent results \citep{khomenko2003,lites2007,lagg2016,martinezgonzalez2016} that found an increasingly magnetic character of the QS. 

These observations were used to infer the magnetic field topology of the QS in the north solar region, at disc centre, and in the West solar region. We focused on the potential differences between the QS magnetism at the PRs and that of the QS at low latitudes as inferred from the disc centre (and west limb) data. Our analysis shows that the magnetic topology at the PR is fully compatible with that inferred at the QS at low latitudes, but projection effects are more relevant closer to the limb and lead to observationally detecting a bi-modal behaviour in the LOS magnetic field topology inferred close to the limbs in the north and west QS regions, which is absent from disc centre.

We analysed the magnetic field topology for the pixels where the splitting between the magnetic components allowed us to avoid possible couplings between the inferred parameters (such as magnetic field strength, inclination, and filling factor). With this selection we were intrinsically biased to strong magnetic fields (above 500-600 G). The disc centre QS region allows optimal observing conditions and a more direct analysis of the results than the limb datasets. At this disc position, we found that the topology of the magnetic fields agrees with the well known canopy picture. These structures are characterised by vertical fields surrounded by a radial azimuth distribution. The same magnetic field topology is also found in the limb datasets. The change from LOS magnetic field parameters to the LRF parameters shows, in addition, that these fields are scattered in patches of a few arcseconds squared \citep{tsuneta2008b,ito2010,jin2011,shiota2012,kaithakkal2013}.

The observed magnetic field inclination and azimuth distributions close to the limb show another population of fields that is not evident at the disc centre. The interpretation of these fields as due to individual magnetic fields gives rise to very singular orientations. When the $180^{\circ}$ solution is adopted, there are two possible magnetic field topologies. First, these fields can present a horizontal configuration. This magnetic field configuration is considered very unlikely because for the observed field strengths (from 600 G to 1000 G), theory predicts buoyant forces that cause these fields to tend to vertical configurations. We also found no horizontal fields at disc centre for the same magnetic field strengths. Second, these magnetic fields can be retrieved with mid-inclinations ($\sim$40$^{\circ}$) and aligned with the disc centre direction (defined by the union of visible disc centre point and the pixel considered). These solutions, either horizontal or with a very particular orientation, and because this occurs in both limb datasets made us assume that projection effects play a relevant role in these inferred magnitudes.

In order to further explore this possibility, we determined whether there is any magnetic structure at disc centre that when modified by projection effects, can reproduce the observed LOS magnetic field inclination and azimuth distributions. We showed that spatially unresolved magnetic loops close to the limb (as foreshortening strongly reduces the spatial resolution close to the limb) can explain the appearance of this LOS field topology, which is not observed at disc centre. We also showed that the QS magnetic topology at the north PR and at low latitudes is fully compatible. This means that the PR magnetic topology is similar to that of the QS at low latitudes.

Strong vertical magnetic fields concentrated in a few arcseconds squared at the PRs were also found by \cite{tsuneta2008b,ito2010,jin2011,kaithakkal2013} and \cite{quinteronoda2016}. By comparing the magnetic structures at high and at low latitudes, we found that both regions show a compatible topological behaviour. This result is in agreement with that by \cite{ito2010} and \cite{jin2011}, who compared the magnetic field topology at both disc positions. \cite{ito2010} also found that the vertical magnetic field structures at the PR carry more magnetic flux than the structures at low latitudes. \cite{shiota2012} found that the number of these higher flux concentrations vary with cycle phase. We found similar flux concentrations at both limbs, which is consistent with these previous works.

Using the same spectral region as used in this study, \cite{blancorodriguez2010} inferred magnetic field strengths ranging from 150 to 1500 G. They also found that magnetic fields above 600 G tend to be vertical and have filling factors in between 0.05 and 0.3. Taking into account our threshold (600 G), the results we reported partially agree with theirs. We also determined magnetic fields that range from 600 to 1500 G and tend to be vertical and with similar filling factors. However, we reported another magnetic feature at high heliocentric angles: unresolved loops at these disc positions, which were not observed by \cite{blancorodriguez2010}. A possible explanation for this difference is that they studied QS faculae, which, according to \cite{kaithakkal2013}, are most likely for the largest magnetic field concentrations, which are associated with vertical fields. Here, we did not restrict ourselves to QS faculae, but analysed all the features associated with polarimetric signals with an inferred magnetic field higher than 600 G.

Most of the topological studies of the polar magnetism \citep{tsuneta2008b, ito2010, blancorodriguez2010, jin2011,kaithakkal2013, quinteronoda2016} have been performed close to solar activity minimum. During this phase of the solar activity cycle, PRs exhibit their strongest magnetic signals. We have studied the magnetic topology at PRs close to activity maximum, when PRs reverse their dominant polarity. Very few studies have been reported that describe this \citep{shiota2012}. We conclude that the magnetic field topology in the north region is the similar to that of the QS at low latitudes. We showed that the observed polarimetric signals are compatible with small-scale magnetic loops at large heliocentric angles. This detection of ubiquitous magnetic loops at the QS, together with the increasing evidence of the influence of these structures in higher layers of the solar atmosphere \citep{martinezgonzalez2009,gomory2013}, confirm the potential role of these QS structures in the energetics of the solar atmosphere \citep{wiegelmann2010}.

\begin{acknowledgements}
Financial support by the Spanish Ministry of Economy and Competitiveness and the European FEDER Fund through projects AYA2014-60476-P (Solar magnetometry in the era of large solar telescopes) and AYA2014-60833-P (Spectropolarimetry: a window to stellar magnetism) is gratefully acknowledged. The German VTT is operated by the Kiepenheuer-Institut f\"ur Sonnenphysik at the Spanish Observatorio del Teide of the Instituto de Astrof\'{\i}sica de Canarias (IAC). This paper made use of the IAC Supercomputing facility HTCondor (\url{http://research.cs.wisc.edu/htcondor/}). NSO/Kitt Peak FTS data used here were produced by NSF/NOAO.
\end{acknowledgements}

%
   \bibliographystyle{aa} 
   \bibliography{biblio.bib} 
%

\end{document}